\documentclass[useAMS,usenatbib]{mn2e}

\usepackage{graphicx}

\title[The Far Outer Disc of M31] {The Star Formation History and Dust
  Content in the Far Outer Disc of M31\thanks{Based on observations
  made with the NASA/ESA {\it Hubble Space Telescope}, obtained at
  the Space Telescope Science Institute, which is operated by the
  Association of Universities for Research in Astronomy, Inc., under
  NASA contract NAS5-26555. These observations are associated with
  program 9458.}}
\author[Edouard J. Bernard et al.]{Edouard J.
  Bernard$^{1}$\thanks{E-mail: ejb@roe.ac.uk}, Annette M. N.
  Ferguson$^{1}$, Michael K. Barker$^{1}$, \newauthor Sebastian L.
  Hidalgo$^{2,3}$, Rodrigo A. Ibata$^{4}$, Michael J. Irwin$^{5}$,
  Geraint F. Lewis$^{6}$, \newauthor Alan W. McConnachie$^{7}$,
  Matteo Monelli$^{2,3}$, Scott C. Chapman$^{5}$\\
  $^{1}$SUPA, Institute for Astronomy, University of Edinburgh, Royal
  Observatory, Blackford Hill, Edinburgh, EH9 3HJ UK. \\
  $^{2}$Instituto de Astrof\'isica de Canarias, V\'ia L\'actea s/n,
  E-38200 La Laguna, Tenerife, Spain.\\
  $^{3}$Departamento de Astrof\'isica, Universidad de La Laguna,
  Tenerife, Spain.\\
  $^{4}$Observatoire de Strasbourg, 11, rue de l'Universit\'e,
  F-67000 Strasbourg, France. \\
  $^{5}$Institute of Astronomy, Madingley Road, Cambridge, CB3 0HA UK. \\
  $^{6}$Institute of Astronomy, School of Physics, University of Sydney,
  NSW 2006, Australia. \\
  $^{7}$NRC Herzberg Institute for Astrophysics, 5071 West Saanich
  Road, Victoria, V9E 2E7, British Columbia, Canada. }

\begin{document}

\date{Accepted 2011 November 18.  Received 2011 November 18; in original form 2011 September 9}

\pagerange{\pageref{firstpage}--\pageref{lastpage}} \pubyear{2011}

\maketitle

\label{firstpage}

\begin{abstract}
  We present a detailed analysis of two fields located 26~kpc ($\sim5$
  radial scalelengths) from the centre of M31 along the south-west
  semimajor axis of the disc. One field samples the major axis
  populations -- the Outer Disc field -- while the other is offset by
  $\sim$18$\arcmin$ and samples the warp in the stellar disc -- the
  Warp field. The color-magnitude diagrams (CMDs) based on {\it Hubble
  Space Telescope} Advanced Camera for Surveys imaging reach old
  main-sequence turn-offs ($\sim 12.5$~Gyr). We apply the CMD-fitting
  technique to the Warp field to reconstruct the star formation
  history (SFH). We find that after undergoing roughly constant star
  formation until about 4.5~Gyr ago, there was a rapid decline in
  activity and then a $\sim$1.5 Gyr lull, followed by a strong burst
  lasting 1.5~Gyr and responsible for 25\% of the total stellar mass
  in this field. This burst appears to be accompanied by a decline in
  global metallicity which could be a signature of the inflow of
  metal-poor gas.  The onset of the burst ($\sim$3~Gyr ago)
  corresponds to the last close passage of M31 and M33 as predicted by
  detailed N-body modelling, and may have been triggered by this
  event.  We reprocess the deep M33 outer disc field data of
  \citet{bar11} in order to compare consistently-derived SFHs.  This
  reveals a similar duration burst that is exactly coeval with that
  seen in the M31 Warp field, lending further support to the
  interaction hypothesis.  We reliably trace star formation as far
  back as 12-13 Gyr ago in the outer disc of M31 while the onset of
  star formation occured about 2~Gyr later in M33, with median
  stellar ages of 7.5~Gyr and 4.5~Gyr, respectively. The complex SFHs
  derived, as well as the smoothly-varying age-metallicity relations,
  suggest that the stellar populations observed in the far outer
  discs of both galaxies have largely formed in situ rather than
  migrated from smaller galactocentric radii. The strong
  differential reddening affecting the CMD of the Outer Disc field
  prevents derivation of the SFH using the same method.  Instead, we
  quantify this reddening and find that the fine-scale distribution of
  dust precisely follows that of the \mbox{H\,{\sc i}} gas. This
  indicates that the outer \mbox{H\,{\sc i}} disc of M31 contains a
  substantial amount of dust and therefore suggests significant metal
  enrichment in these parts, consistent with inferences from our CMD
  analysis.
\end{abstract}

\begin{keywords}
galaxies: individual (M31) --
galaxies: interactions --
galaxies: stellar content --
ISM: dust, extinction --
Local Group --
stars: variables: RR Lyrae
\end{keywords}

\begin{figure}          
\includegraphics[width=8.5cm]{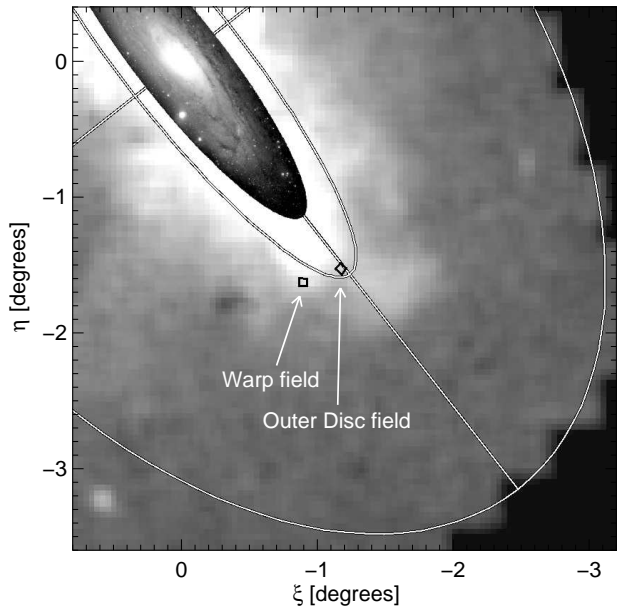}
\caption{Location of the HST fields (small black squares),
 overplotted on the surface density of RGB stars from the INT/WFC
 survey of \citet{irw05} in standard coordinate projection. An
 optical image of the inner disc has been overlaid.  The inner
 ellipse has a semimajor axis of 2$\degr$ (27~kpc) and represents an
 inclined disc with $i$ = 77\fdg 5 and position angle of 38\fdg 1.
 The outer ellipse, of semimajor axis length 4$\degr$ (55~kpc),
 roughly indicates the spatial extent of the INT survey.
 \label{fig:1}}
\end{figure}

\section{Introduction}\label{sec:1}

Disc-dominated galaxies account for a sizeable fraction of the stellar
mass in the Universe \citep[e.g.,][]{gal08,van11} and thus
understanding their formation and evolution is of prime importance. In
the classical picture, a disc galaxy forms from the collapse of a
rotating gaseous protogalaxy within the gravitational potential well
of a dark matter halo. As the gas cools and settles into a
rotationally-supported disc, it can fragment and undergo star
formation \citep[e.g.,][]{fal80}. Modern-day studies place these ideas
within a cosmological $\Lambda$ cold dark matter context and use ever
more advanced simulations to follow the aquisition of mass through
mergers with other systems, as well as from the smooth accretion of
intergalactic gas.

Despite the unprecedented numerical resolution and the impressive
range of gas physics incorporated, simulations have had difficulties
successfully reproducing late-type massive spiral galaxies such as the
Milky Way and M31 \citep[e.g.,][]{gov09,age11} and are only now
starting to achieve it \citep{gue11}.  A key issue is the so-called
`angular momentum problem' which leads to overly small,
centrally-concentrated stellar discs dominated by large bulges.
Proposed solutions generally invoke various forms and degrees of
stellar feedback that suppress star formation at early epochs thereby
allowing extended discs to form, albeit at relatively recent times
(z$<$1). Due to the expected inside-out growth of the disc, such
models require young mean stellar ages at large radii and interesting
constraints arise from observations of old and intermediate age stars
in these parts \citep[e.g.,][]{fer01,wys08}.

Once in place, various processes may influence the star formation and
chemical enrichment history of galactic discs. Near or long-range
interactions with neighbouring galaxies can redistribute gas and stars
and trigger bursts of star formation in which significant fractions of
the total stellar mass can form \citep[e.g.,][]{dim08,won11}.  Various
authors have also shown how stars can undergo large radial excursions
due to scattering off recurrent transient spirals and other features
\citep[e.g.,][]{sel02,ros08a,min11} as well as to perturbations caused
by satellite accretion \citep{bir11}. The primary effect is to
efficiently mix stars over kiloparsec scales within discs, the
consequences of which could be quite profound. For example, if
considerable radial migration takes place in discs, it would pose a
severe problem for interpreting analyses of the resolved fossil record
since the current location of an old star may be far removed from its
actual birth place \citep{ros08b}.  However, as of yet, there is little
direct and unambiguous observational evidence for radial migration
\citep[e.g.,][]{haywood08}.

\begin{table}
 \centering
 \begin{minipage}{85mm}
  \caption{Summary of Observations\label{tab1}}
  \begin{tabular}{@{}lcc@{}}
  \hline
  Field                                                                           &  Warp                             & Outer Disc           \\
  \hline
  R.A. (J2000)                                                                    &  ~~00:38:05.1                     & ~~00:36:38.3         \\
  Decl. (J2000)                                                                   & +39:37:55                      & +39:43:26    \\
  $R_{proj}$\footnote{Projected radial distance.} (kpc)                            &  25.4                             & 26.4                 \\
  $R_{disk}$\footnote{Deprojected radial distances are calculated assuming
    (m$-$M)$_0$=24.47 \citep{hol98} and an inclined disc with P.A. = 38\fdg 1 \citep{fer02} and $i$ = 77\fdg 5 \citep{ma97}.} (kpc) & 31.5                        & 26.4                 \\
  E(B$-$V)\footnote{Values from \citet{sch98}.}                                   & 0.054                             & 0.055                \\
  Dates                                                                            & 2003 Jun 9$-$12                   & 2003 Aug 13$-$16     \\
  t$_{F606W}$\footnote{Individual and total exposure times in the F606W band.} (s) & 8$\times$1310=10480               & 8$\times$1250=10000  \\
  t$_{F814W}$\footnote{Individual and total exposure times in the F814W band.} (s) & 4$\times$1250+8$\times$1330=15640 & 4$\times$1200+8$\times$1280=15040 \\
  \hline
\end{tabular}
\end{minipage}
\end{table}

\begin{figure*}          
\begin{center}
\includegraphics[width=12.5cm]{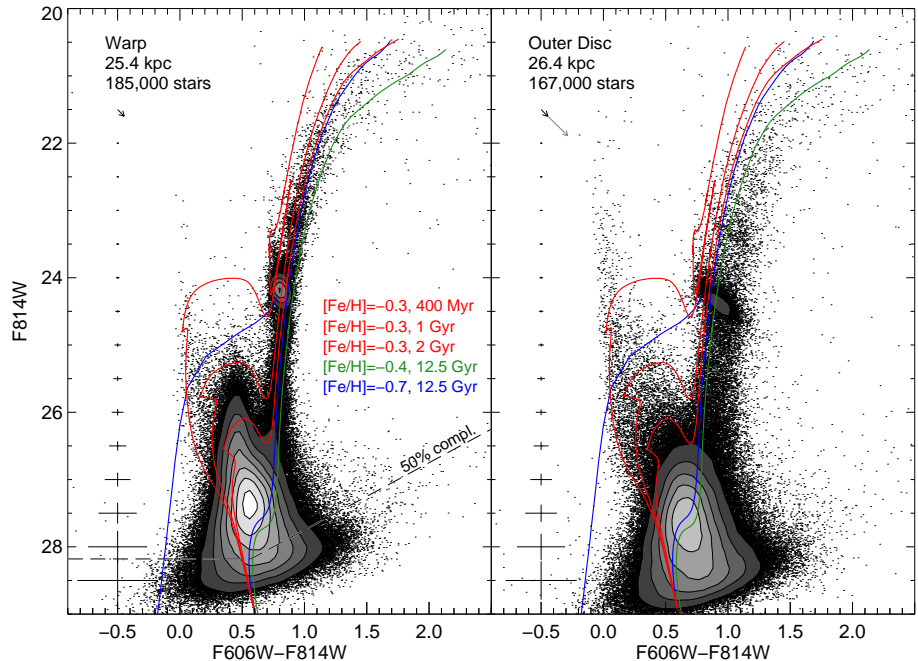}
\caption{Color-magnitude diagrams for the Warp (left) and Outer Disc
 (right) fields, where selected isochrones and a ZAHB from the BaSTI
 library \citep{pie04} are overlaid. The error bars show the mean
 photometric errors as a function of magnitude. In each panel, the projected
 radial distances and the number of stars in the CMDs are indicated.
 The contour levels correspond to [30, 55, 80, 105, 130, 155,
 180]$\times 10^3 {\rm \ stars \ mag}^{-2}$. The 50\% completeness
 limit from the artificial stars tests (see Section~\ref{sec:2.4}) is
 also shown by the dashed line in the left panel.  The arrows in the
 top left of each panel indicate the amplitude of reddening
 \citep[black arrows, from][]{sch98} and differential reddening (gray
 arrow, see Section~\ref{sec:5}). Note the large spread in the
 features of the Outer Disc field due to strong differential
 reddening.
 \label{fig:2}}
\end{center}
\end{figure*}

The best place to test ideas about disc formation and evolution is in
the nearest external systems where stellar populations with a broad
range of age, including those formed at cosmologically-interesting
epochs (z$\sim$1--2), can be resolved into individual stars. There has
been remarkable recent progress in this field, thanks largely to the
ANGST and LCID collaborations which have obtained high-quality Hubble
Space Telescope ({\it HST}) observations of resolved populations in a
variety of nearby galaxies \citep[e.g.,][]{ber09, dalc09,
wil09a,mon10, hid11}.  Of particular importance are studies of
populations in the outermost regions of spiral discs: although these
regions contain only a minor fraction of the total disc mass, they are
where many evolutionary models diverge the most in their predictions
and are therefore most easily tested.  Interestingly, while the star
formation histories (SFHs) reported thus far show evidence of
inside-out disc growth \citep[e.g.,][]{wil09b,wil10,gog10,bar11}, they
also indicate that a significant fraction of the mass in discs was in
place by z$\sim$1--2.

Over the past few decades, HST color-magnitude diagrams (CMDs) of the
outer regions of the our largest Local Group neighbours, M31 and M33,
have been obtained and analysed in the context of their SFHs. Some of
these observations have been limited in depth \citep{wil02} while
others have had their interpretations complicated by the presence of
multiple structural components and/or halo substructure
\citep{bro06,bar07,ric08, ric10}.  Studies of fields sampling the
entire M33 disc have revealed a clear detection of an inverse age
gradient which reverses across the disc truncation
\citep{wil09b,bar11}. During Cycle 11, 47 {\it HST} orbits were
dedicated to observe eleven fields in the outskirts of M31 with the
Advanced Camera for Surveys (ACS; P.ID. 9458, P.I.: A.  Ferguson).
Two outer disc fields were observed for ten orbits each, while three
orbits were secured for each of the remaining halo substructure
fields.  The analysis of the stellar populations of the substructure
based on this dataset has been presented in an earlier series of
papers.  \citet{fer05} explored the stellar populations associated
with five prominent stellar overdensities in the halo of M31, and
found large-scale population inhomogeneities in terms of age and
metallicity.  A more in-depth analysis of one of these fields, the G1
Clump, placed constraints on the SFH of this region and noted the
similarity to the M31 outer disc population \citep{far07}.  Finally,
\citet{ric08} combined this dataset with other deep archival datasets
in order to explore the nature and origin of the substructure. The
global picture that emerges from their homogeneous analysis of 14 HST
CMDs is that the substructure can be largely explained as either
stripped material from the Giant Stellar Stream progenitor
\citep{iba01} or material torn off from the thin disc through
disruption.

Here, we focus on the analysis of the two deep disc pointings sampling
the Warp and Outer Disc. While the Warp field was previously presented
in \citet{ric08}, it was analysed at a only fraction of its total
depth (three out of the ten orbits) and the observations presented
here are roughly a full magnitude deeper.
The paper is structured as follows: in Section~\ref{sec:2}, we
describe the observations and data reduction, and present the
resulting CMDs. The CMD-fitting method used to recover the SFH is
described in Section~\ref{sec:4}, along with the results for the Warp
field in M31 and the S1 field in M33.  In Section~\ref{sec:5}, the
differential reddening affecting the CMD of the Outer Disc field is
measured, and used to constrain the fine-scale distribution of dust in
the field.  The multi-epoch nature of the datasets allowed us to
search for variable stars in both fields; they are briefly described
in Section~\ref{sec:3}.  Our interpretation of the results is
discussed in Section~\ref{sec:6} and a summary of the main results is
given in Section~\ref{sec:7}.

\section{The Data}\label{sec:2}

\subsection{Observations}\label{sec:2.1}

\begin{figure}          
\includegraphics[width=8.5cm]{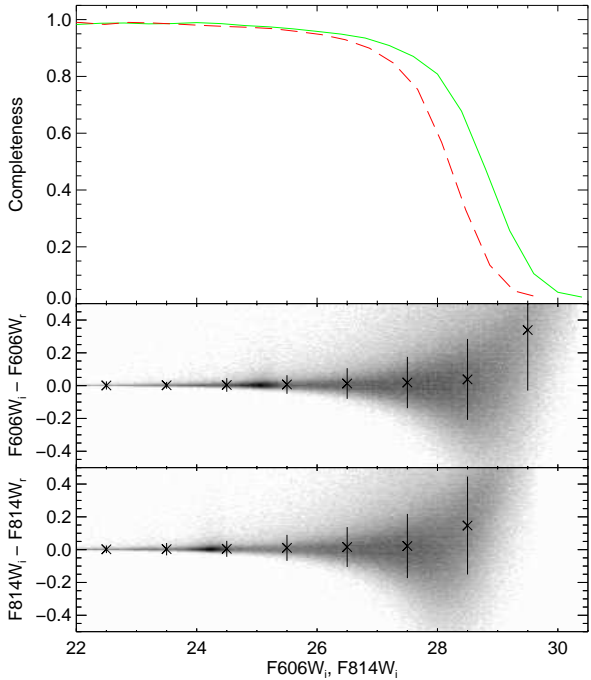}
\caption{Results from the artificial star tests for the Warp field.
 Top: Completeness in the F606W (green, solid line) and F814W (red, dashed line)
 bands as a function of injected magnitude. Middle (bottom) panel: residuals
 between injected and recovered F606W (F814W) magnitudes as a function of
 injected magnitude. The crosses and lines indicate the median and standard
 deviation of the residuals calculated in 1~mag bins.
 \label{fig:3}}
\end{figure}

The two disc fields have been observed with the Advanced Camera for
Surveys onboard the {\it HST}. Their location was chosen so that the
on- and off-axis stellar populations could be sampled and compared.
Throughout this paper, we assume a distance of 783~kpc, corresponding
to (m$-$M)$_0$=24.47 \citep{hol98}, and an inclined disc with P.A. =
38\fdg 1 \citep{fer02} and $i$ = 77\fdg 5 \citep{ma97}.  Located
$\sim$18$\arcmin$ apart, the $`$Outer Disc' field samples the outer HI
disc along the major axis at projected (deprojected) radius of $\sim
26$ (26)~kpc, while the $`$Warp' field samples the stellar disc at a
projected (deprojected) radius of $\sim 25$ (31)~kpc, where it
strongly bends away from the major axis. Note that the deprojected
radii, which correspond to circular radii within the disc plane, are
only valid in the case of a planar disc, which is clearly not true in
the outer regions of M31.  Assuming a dust-free exponential disc
scalelength of $\rm{R}_d = 5.3 \pm 0.5\ \rm{kpc}$ \citep{cou11}, these
radii correspond to $\sim $5--6 scalelengths.

The fields are shown in Figure~\ref{fig:1}, overplotted on a map of the
surface density of RGB stars from the INT/WFC survey of \citet{irw05}.
For each field, 8 and 12 images were secured in the F606W and F814W
bands during about four consecutive days. Exposure times of
$\sim$1,300 seconds per exposure lead to a total of about 10,000 and
15,000 seconds in F606W and F814W, respectively. A detailed summary of
observations including coordinates, foreground color excess, projected
and deprojected radial distances, and exposure times is given in Table
~\ref{tab1}.

\subsection{Photometry and Calibration}\label{sec:2.2}

The pipeline processed, non-drizzled (`{\tt FLT}') data products were
retrieved from the archive and each exposure was split into two
individual images corresponding to the two ACS chips.  Additional
processing to correct for intra-field variations of the pixel areas on
the sky and to flag bad pixels was done using the pixel area maps and
data quality masks, respectively. Stellar photometry was carried out with
the standard DAOPHOT/ALLSTAR/ALLFRAME suite of programs \citep{ste94} as
follows.  We performed a first source detection at the 3-$\sigma$
level on the individual images, which was used as input for aperture
photometry.  From this catalog, 300 bright, non-saturated stars per
image were initially selected as potential PSF stars. An automatic
rejection based on the shape parameters was used to clean the lists,
followed by a visual inspection of all the stars to remove the
remaining unreliable stars. We ended up with clean lists containing at
least 200 good PSF stars per image.  Modelling of the empirical PSF
with a radius of 10 pixels (0.5$\arcsec$) was done iteratively with
DAOPHOT: the clean lists were used to remove all the stars from the
images except PSF stars, so that accurate PSFs could be created from
non-crowded stars. At each iteration, the PSF was modeled more
accurately and thus the neighbouring stars removed better. Every 3 to
5 iterations, the degree of PSF variability across the image was also
increased, from constant to linear, then quadratically variable.

The following step consisted of profile-fitting photometry on
individual images using ALLSTAR with the empirical PSFs previously
created. The resulting catalogs were used to calculate the geometric
transformations between the individual images using
DAOMATCH/DAOMASTER. From these transformations, it was possible to
create a median, master image by stacking all the images from both
bands using the stand-alone program MONTAGE2. This master image, clean
of cosmic rays and bad columns and much deeper than any individual
frame, was used to create the input star list for ALLFRAME by
performing a second source detection.

The output of ALLFRAME consists of a catalog of PSF photometry for
each image.  A robust mean magnitude was obtained for each star by
combining these files with DAOMASTER, keeping only the stars for which
the PSF fitting converged in at least (N$_i$/2)+1 images per band,
where N$_i$ represents the total number of images per band. This
constraint greatly limits the number of false detections due to noise
at faint magnitudes.

\begin{figure*}          
\begin{center}
\includegraphics[width=12.5cm]{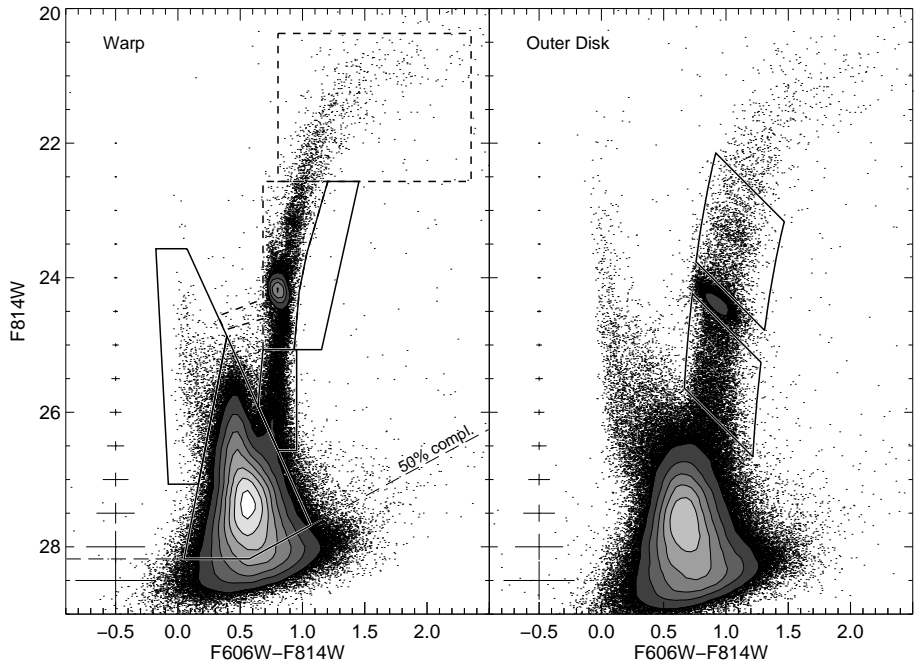}
\caption{Color-magnitude diagrams for the Warp and Outer Disc fields
 showing the location of the regions used for the CMD-fitting (left)
 and for estimating the differential reddening (right, see
 Section~\ref{sec:5}). In the left panel, the dashed boxes
 indicate the location of the two additional bundles used to test
 our SFH calculation (see Figure~\ref{fig:a8}).
 \label{fig:3b}}
\end{center}
\end{figure*}

The final photometry was calibrated to the VEGAMAG system following
the prescriptions of \citet{sir05}, with the revised ACS
zeropoints\footnote{posted 2009 May 19 on the HST/ACS webpage:
  http://www.stsci.edu/hst/acs/analysis/zeropoints.}. Note that the
CMD-fitting technique used in this paper minimizes the impact of the
uncertainties in distance, mean reddening, and photometric zero-points
on the solutions by shifting the observed CMD with respect to the
artificial CMD; therefore, any small errors in these quantities do not
affect the results.

\subsection{Color-Magnitude Diagrams}\label{sec:2.3}

Figure~\ref{fig:2} shows the CMDs for the Warp ({\it left}) and Outer
Disc ({\it right}) fields. These were cleaned of non-stellar objects
using the photometric parameters given by ALLFRAME, namely
$\sigma_{606,814} \leq$~0.3 and $-$0.3~$\leq$~{\tt SHARP}~$\leq$~0.3.
Selected isochrones from the BaSTI library \citep{pie04} are shown,
both to indicate the location of the oldest main-sequence turn-offs
(MSTO) and as a guide for comparing the two CMDs.  As already noted by
\citet{ric08}, the horizontal-branch (HB) of the Warp field is very
sparsely populated. A theoretical zero-age horizontal-branch (ZAHB)
for [Fe/H]=$-$0.7 from the same stellar evolution library is
overplotted to show its expected location. The reddening \citep[black
arrows, from][]{sch98} and differential reddening (gray arrow, see
Section~\ref{sec:5}) at these locations are also shown in the top left
of each panel.

Despite their apparent proximity to each other in the south-western
part of the disc, the CMDs of the two fields present notable
differences.  While the main sequence (MS) of the Warp field contains
very few stars younger than $\sim250-300$~Myr, the Outer Disc field
harbors stars as young as a few tens of million years. On the other
hand, a plume of stars at 0.3~$\leq$~F606W$-$F814W~$\leq$~0.6 and
F814W~$\ga$~25.5, indicative of strong star formation until about
1~Gyr ago, is far more prominent on the CMD of the Warp field.

Concerning the evolved stages of evolution, both fields exhibit
well-populated red giant branch (RGB) and red clump (RC) features.
There are also hints of AGB bumps at F814W$\sim$23.2. The Outer Disc
RGB has a much larger color dispersion than that of the Warp, and the
fact this spread is along the direction of the reddening vector,
especially apparent from the morphology of the red clump (RC),
indicates that the field is strongly affected by differential
reddening. We will discuss and quantify this reddening in
Section~\ref{sec:5}.

Finally, the photometry of the Warp field is about 0.3~mag shallower
than that of the Outer Disc field. This is mainly due to the higher
background at the time of observation.

\subsection{Artificial Star Tests}\label{sec:2.4}

In order to retrieve an accurate SFH from a CMD, a good knowledge of
the external factors affecting the photometry is necessary. This is
usually done through extensive artificial star tests, whereby a large
number of artificial stars are added to the original images, and the
photometry is repeated exactly the same way as for the real stars. The
comparison between the injected and recovered magnitudes reveals the
biases due to observational effects, such as signal to noise
limitations, stellar crowding, and blending, as well as quantifies the
completeness of the photometry.

Unfortunately, these tests cannot quantify the effect of differential
reddening, as observed in the Outer Disc field and described in the
previous section. In this particular case, the variations are so
strong that the CMD-fitting technique used to recover the SFH
described in Section~\ref{sec:4} would be in vain. Hence, the
artificial star tests were only performed on the images of the Warp
field.

The list of stars added to the images was produced by IAC-star
\citep{apa04}.  It consists of a synthetic CMD with very wide ranges
of ages (0--15~Gyr) and metallicities (Z=0.0001--0.02) in order to
cover larger ranges of color and magnitude than for the observed
stars. The stars were added on a triangular grid, separated by
20~pixels (twice the PSF radius), to avoid overlap between the
artificial stars and optimally sample the whole field of view.
Twenty iterations per image, each with 24,395 artificial stars, were
necessary to obtain a sufficient sampling of the parameter space. In
total, almost a million unique artificial stars were added to the
images for these tests.

Figure~\ref{fig:3} presents the results of the artificial star tests.
The top panel shows the completeness as a function of magnitude for
each band. For the F606W and F814W bands, the photometry is 90\%
complete down to 27.29 and 26.85 mag, respectively, but drops to 50\%
at F606W=28.75 and F814W=28.18. As shown in Figure \ref{fig:2}, this
latter limit roughly corresponds to the magnitude of a 12.5~Gyr old
MSTO.  The middle and bottom panels of Figure~\ref{fig:3} compare the
injected and recovered magnitudes of the artificial stars in each
band, as a function of injected magnitudes. The offset toward
positive residuals at fainter magnitudes is a consequence of crowding
in these fields, which causes stars to be recovered brighter than the
input magnitudes.

\begin{figure}          
\includegraphics[width=8cm]{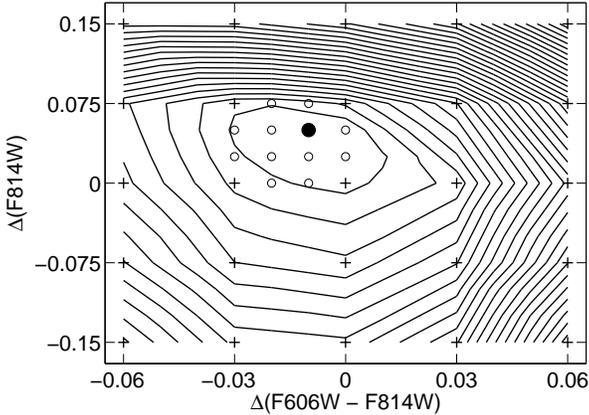}
\caption{Summary of the 888 calculated solutions in the color-magnitude
 plane: 24 solutions were calculated at each location indicated by a plus
 symbol or a circle. The contours show the resulting $\chi^2$ distribution in
 steps of 0.5. Our best-fit solution is the average of the 24 solutions
 indicated by the filled circle.
 \label{fig:7}}
\end{figure}

\section{Quantitative Star Formation History of the Warp Field}\label{sec:4}

\subsection{Method}\label{sec:4.1}

The SFH of the Warp field in M31 was derived through the widely-used
CMD-fitting technique, in which the observed CMD is compared with
synthetic CMDs created using up-to-date models of stellar evolution.
In particular, we adopt the IAC method developed for the LCID Project
\citep[e.g.,][]{mon10}, which uses IAC-star \citep{apa04} to generate
synthetic populations, IAC-pop \citep{apa09} to find the best
solution, and MinnIAC \citep{hid11} to produce the input files for,
and process the output files from, IAC-pop.

The comparison between the observed and synthetic CMDs is done by
simply counting the number of stars in specific areas -- referred to
as $`$bundles' \citep{apa09} -- of the CMDs.  These areas are chosen
on the basis of reliability through considering observational effects
(e.g., signal-to-noise ratio, stellar density) as well as theoretical
uncertainties in stellar evolution models. To limit the effects of the
former, we only consider stars on the CMD brighter than the 50\%
completeness level in both bands.  We also exclude stars on the red
giant branch
brighter than F814W$\sim$25 (i.e., one magnitude below the RC). The
choice to exclude the latter is mainly motivated by the confusion
between the various phases of stellar evolution in this part of the
CMD -- RGB, asymptotic giant branch, RC, red HB, and red supergiant
branch -- as well as the uncertainties in the theoretical models of
these advanced evolutionary stages \citep[see,
e.g.,][]{ski03,gal05,wil09a}.  Indeed, most SFH analyses that are
based on deep MSTO photometry do not include this part of the CMD in
the fitting procedure \citep[e.g.,][]{bro06,mon10,hid11}
although our tests reveal that this does not have a very strong effect
on the resultant solution \citep[see Figure~\ref{fig:a8},
and][]{bar11}. The bundles used for the Warp CMD are shown as solid
lines in the left panel of Figure~\ref{fig:3b}.

The bundles are divided uniformly into small boxes, the size of which
depends on the density of stars and reliability of the stellar
evolution models. The size of these boxes ranges from 0.06$\times$0.04
mag for the bundles covering the MS to 0.3$\times$0.5 for the bundle
located on the red side of the RGB. This yields a total number of
$\sim10^3$ boxes.  Since every box carries the same weight, the much
larger number of boxes on the MS also gives a higher weight to this
well-understood evolutionary phase in the SFH calculations.

The number of observed stars, and artificial stars from each simple
stellar population (SSP), counted in each CMD box serves as the only
input to IAC-pop.  IAC-pop then tries to reproduce the observed CMD as
a linear combination of synthetic CMDs corresponding to SSPs, that is,
populations with small ranges of age and metallicity. No a priori
age--metallicity relation, or constraint thereon, is adopted: IAC-pop
therefore solves for both ages and metallicities simultaneously. The
goodness of the coefficients in the linear combination is measured
through the $\chi^2$ merit function, which IAC-pop minimizes using a
genetic algorithm. These coefficients are directly proportional to the
star formation rate of their corresponding SSPs.

\begin{figure*}          
\begin{center}
\includegraphics[width=15cm]{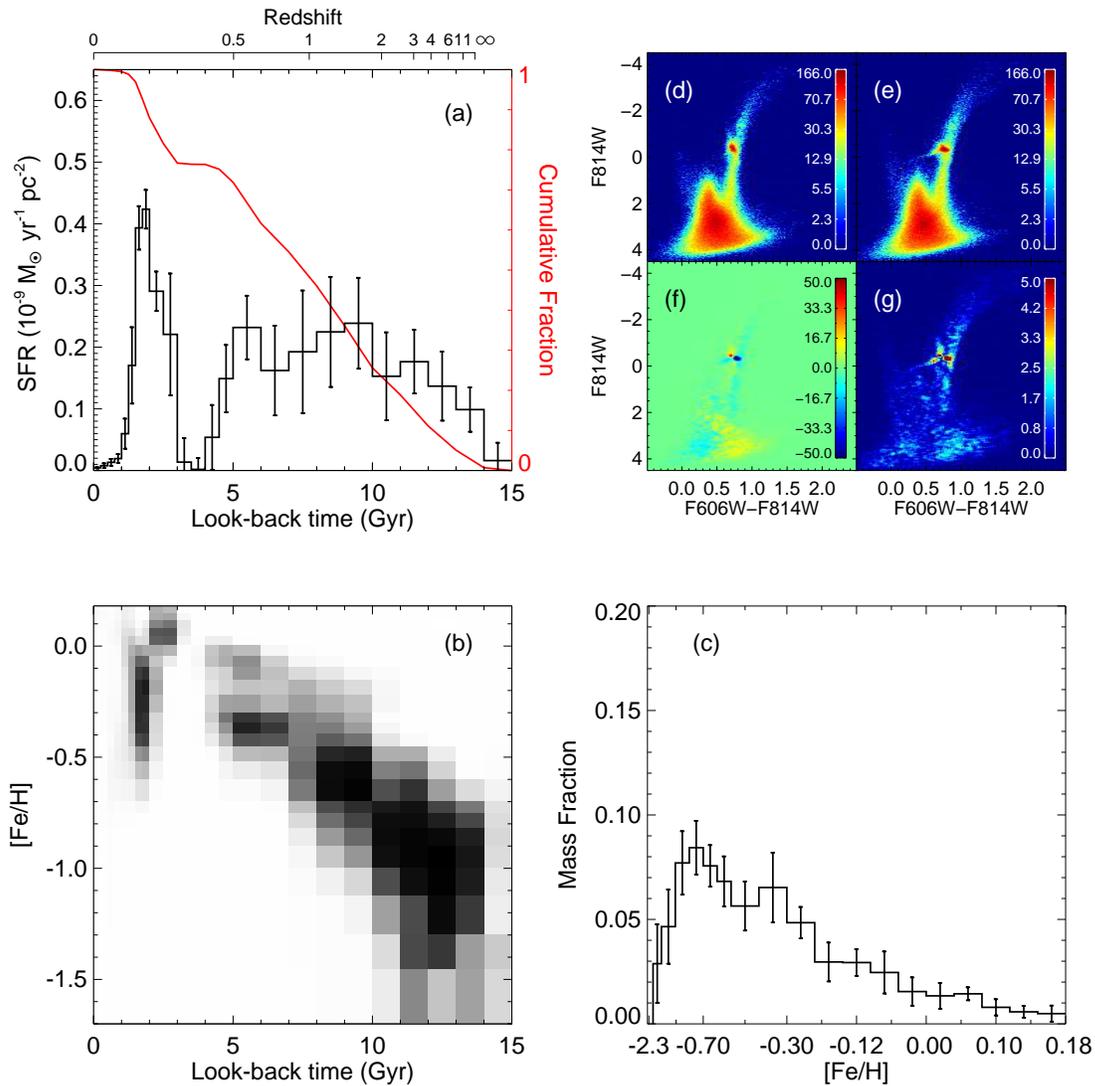}
\caption{Best-fit SFH solution for the Warp field, obtained with the BaSTI
 library. The panels show: (a) the SFR as a function of time, normalised to the
 deprojected area of the ACS field, (b) the age-metallicity relation, (c) the
 metallicity distribution of the mass of stars formed, (d)-(e) the
 Hess diagrams of the observed and best-fit model CMDs, (f) the residuals, and
 (g) the significance of the residuals in Poisson $\sigma$. The cumulative mass
 fraction is shown in red in panel (a). See text for details.
 \label{fig:8}}
\end{center}
\end{figure*}

Age and metallicity are not the only parameters that can influence the
best-fit solution. Uncertainties in the distance, mean reddening,
photometric zero-points and model systematics, as well as the
subjective selection of CMD areas and definition of the SSPs, also
affect the resulting SFH solution.  MinnIAC was specifically developed
to explore this vast parameter space and combine the best solutions at
each gridpoint to obtain a robust, stable overall SFH.  Its two main
purposes are: {\it i)} to automate the production of the input files
to run a large number of IAC-pop processes in order to sample the
whole parameter space, and {\it ii)} to average all the solutions
obtained at a given distance and reddening to reduce the impact of any
specific choices.

We generated a synthetic CMD containing 10$^7$ stars using IAC-star
\citep{apa04} with the following main ingredients. We select the BaSTI
stellar evolution library \citep{pie04}, more specifically the
scaled-solar models with overshooting, within age and metallicity
ranges of 0 to 15~Gyr old and 0.0004$\le \rm{Z} \le$0.03 (i.e.,
$-$1.7$\le$[Fe/H]$\le$0.18 assuming $Z_{\sun}=\rm{0.0198}$;
\citealt{gre93}), respectively. We assume the \citet{kro02} initial
mass function (IMF), which is a broken power law with exponents of
$-$1.3 for stars with masses $<0.5\rm{M}_{\sun}$ and $-$2.3 for higher
masses.  \citet{hid11} carried out extensive tests of the effect of
different IMF slopes for their IC1613 and LGS3 analysis, and found the
best results were obtained with values compatible with the Kroupa IMF.
They did note, however, that the resulting SFH was rather insensitive
to realistic changes of the IMF slope.  In addition, since our data do
not reach the lower mass range, a different choice of IMF slope only
affects the normalization and not the relative star formation rates
\citep[SFRs; see also][]{bro06,bar07}.  Finally, the binary fraction
and mass ratio were set to f=0.5 and q$>$0.5, respectively, which is
in agreement with G and M star surveys of the Milky Way field
population \citep[e.g.,][]{duq91,rei97}. According to the tests
performed by \citet{mon10} on five galaxies, the binary fraction only
marginally affects the derived SFHs.

\begin{figure*}          
\begin{center}
\includegraphics[width=12.5cm]{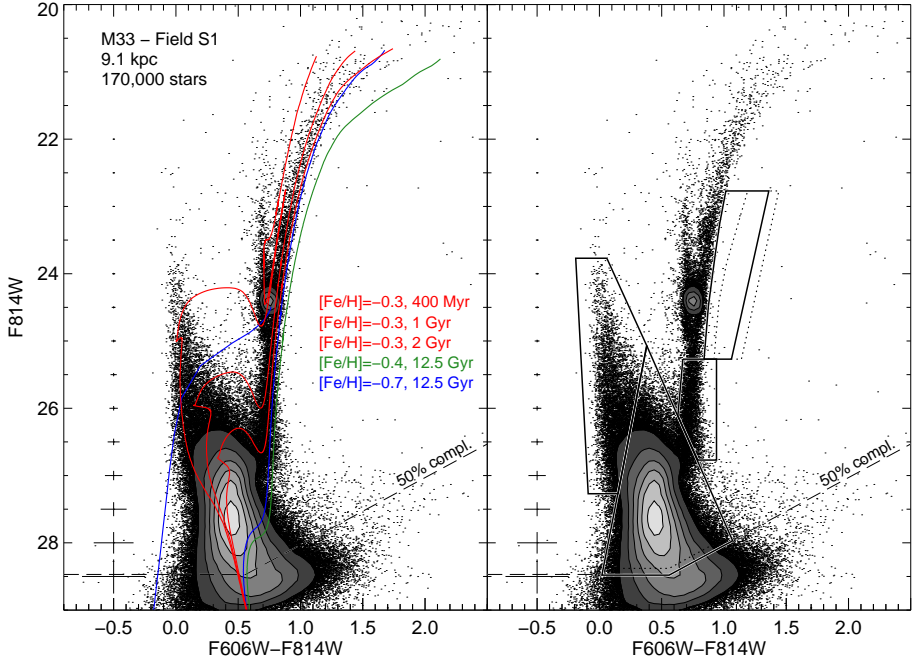}
\caption{Color-magnitude diagram for the reprocessed M33 field S1 from
 \citet{bar11}, where the deprojected radial distance and the number
 of stars in the CMD are indicated. The contour levels, isochrones,
 and ZAHB are the same as in the left panel of Figure~\ref{fig:2} to
 facilitate comparison. The 50\% completeness limit from the
 artificial stars tests is also shown by the dashed line.  Right:
 same as left panel, only showing the location of the four bundles
 used for the SFH calculations. The dotted lines show the bundles
 used for M31, shifted to the distance and reddening of M33.
 \label{fig:8c}}
\end{center}
\end{figure*}

As a first step, the age--metallicity plane was divided in 19 age bins
and 10 metallicity bins, producing the equivalent of 190 SSPs. The
choice of bin sizes ($\Delta$(age): 0.5~Gyr between 0 and 4~Gyr old,
1~Gyr beyond; metallicity boundaries at Z=[0.0004 0.0007 0.0010 0.0020
0.0040 0.0070 0.0100 0.0150 0.0200 0.0250 0.0300]) was made after
testing various combinations. However, as stated above, the exact
choice of these simple populations can affect the solution, so the SFH
was calculated several times, each time slightly shifting the age and
metallicity limits of each bin, as well as the location of the boxes
on the CMDs in which stars are counted. As was done for the LCID
Project, the age and metallicity bins were shifted three times, each
time by an amount equal to 30\% of the bin size, with four different
configurations: (1) moving the age bin toward increasing age at fixed
metallicity, (2) moving the metallicity bin toward increasing
metallicity at fixed age, (3) moving both bins toward increasing
values, and (4) moving toward decreasing age and increasing
metallicity. These twelve different sets of parameterization were used
twice, shifting the boxes a fraction of their size across the CMD.
These 24 individual solutions were then averaged to obtain a more
stable solution. The $\chi^2$ of this average solution is simply the
average of the 24 individual reduced $\chi^2$, which we denote
$\overline{\chi^2}$.

This process is then repeated several times after shifting the
observed CMD with respect to the synthetic CMDs in order to account
for uncertainties in photometric zero-points, distance and mean
reddening. Here we shift the CMD 25~times on a regular 5$\times$5 grid
within the following limits: $-0.06\leq\Delta(F606W-F814W)\leq0.06$
and $-0.15\leq\Delta F814W\leq0.15$, as shown in Figure~\ref{fig:7}.
These limits are roughly twice the uncertainties in reddening and
distance, and correspond to $\Delta \rm{E(B-V)}\sim0.06$ and $\Delta
\rm{D}\sim55$~kpc. Each plus symbol in Figure~\ref{fig:7} represents
one of the nodes of this grid, where 24 individual solutions were
calculated and averaged.  This produced a $\chi^2$--map where the
contours indicate the approximate location of the absolute minimum in
the distance--reddening plane.  A second series of finer shifts around
this minimum was applied to refine its location; these are shown as
the 12 circles in Figure~\ref{fig:7}. Therefore, 888 individual
solutions were calculated in total. The filled circle indicates the
location of the averaged solution with the lowest reduced
$\overline{\chi^2}$, i.e., our best-fit solution SFH, with
$\overline{\chi^2}_{\rm{min}}=3.08$.  We stress that we do not
consider the position of the best $\overline{\chi^2}$ in the
distance--reddening map as a reliable estimate of distance or
reddening since photometric zero points, model uncertainties, and
other hidden systematics will affect its absolute location.

The uncertainties on the SFRs were estimated following the
prescriptions of \citet{hid11}.  The total uncertainties are assumed
to be a combination of the Gaussian uncertainties $\sigma_G$ -- which
include errors on distance, reddening, and photometric zero points, as
well as the effect of sampling in the color--magnitude and
age--metallicity planes -- and Poissonian uncertainties $\sigma_P$,
due to the effect of statistical sampling in the observed CMD.  Given
the large number of solutions calculated after varying the input
parameters, the rms deviation of all the `good' solutions can be
assumed to be a reliable proxy of $\sigma_G$. Defining
$\overline{\chi^2}_{\rm{min}}$ and
$\sigma_{\overline{\chi^2}_{\rm{min}}}$ as the lowest reduced
$\overline{\chi^2}$ and its standard deviation, we consider `good' all
the \emph{individual} solutions with $\chi^2 \leq
\overline{\chi^2}_{\rm{min}} + \sigma_{\overline{\chi^2}_{\rm{min}}}$,
regardless of their location on the $\chi^2$--map.

\begin{figure*}          
\begin{center}
\includegraphics[width=15cm]{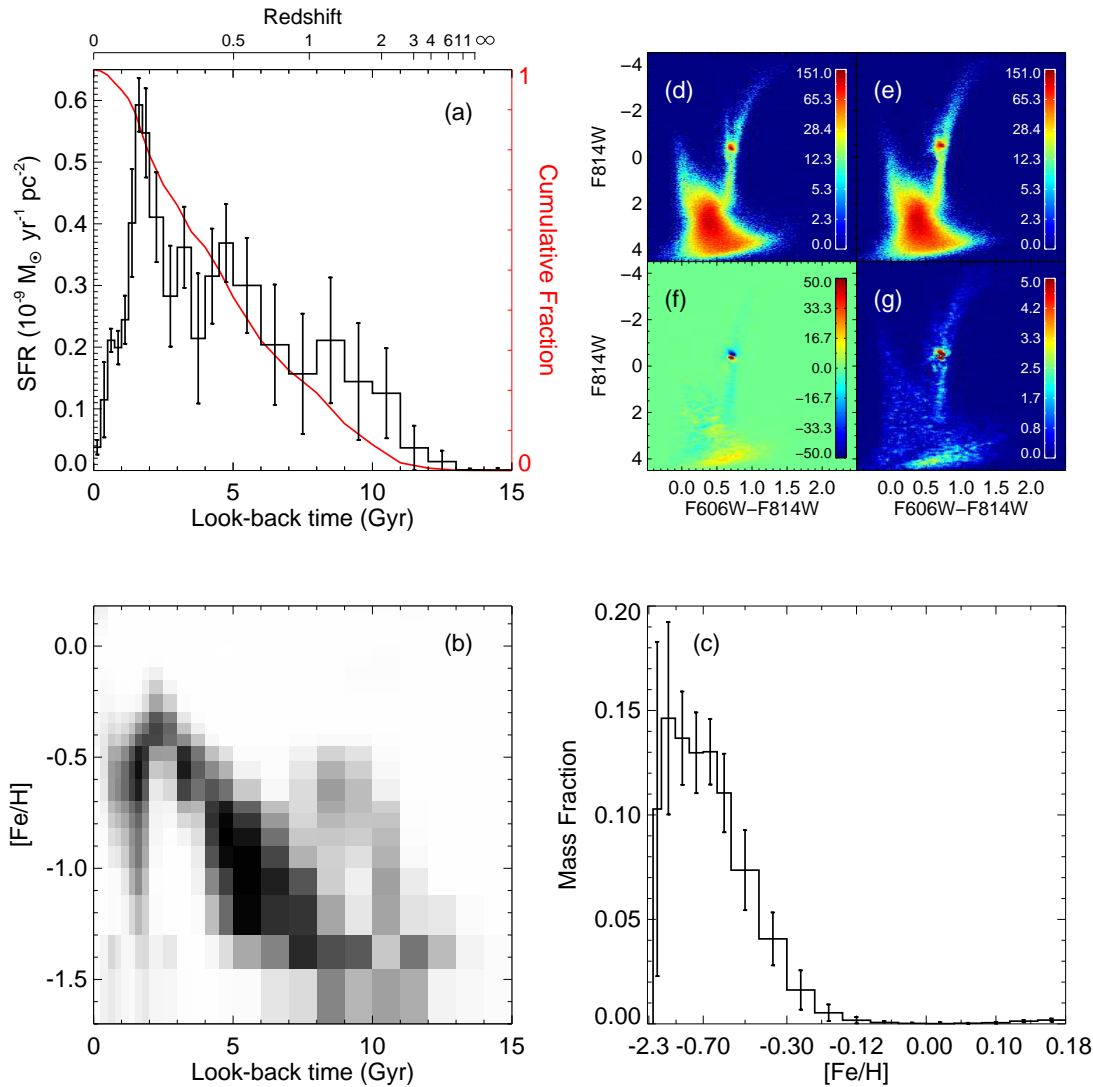}
\caption{Same as Figure~\ref{fig:8}, but for the reprocessed M33 field S1 of
 \citet{bar11}.
\label{fig:8b}}
\end{center}
\end{figure*}

The Poissonian errors are determined by varying the number of stars in
each box of the observed CMD according to Poisson statistics, with all
the other parameters held fixed, and recalculating the solution. This
process is repeated 20 times, and the rms dispersion of these
solutions is taken as $\sigma_P$. The total uncertainty is the sum in
quadrature of $\sigma_G$ and $\sigma_P$.

To assess the significance and reliability of the SFH solution and to
understand the possible systematics, we repeated the calculation of
the Warp SFH several times either: i) using the Padova stellar
evolution library \citep{gir00}, ii) constraining the range of
metallicities for young ages ($<$3.5~Gyr), or iii) including the
bright end of the CMD; as well as tried to recover the SFH of a series
of artificial CMDs for which the input SFHs were known. The
calculations were done strictly following the method described in this
section.  The results proved very reassuring and are presented and
discussed in the Appendix.

\subsection{Results}

Our best-fit solution SFH is presented in Figure~\ref{fig:8}. The
different panels show: the evolution of the SFR (a) and metallicity
(b) as a function of look-back time; (c) the metallicity distribution
of the mass of stars formed; (d)-(e) the Hess diagrams of the observed
and best-fit model CMDs; (f) the residual differences between the two,
in the sense observed$-$model; and (g) the absolute residual
difference normalized by Poisson statistics in each bin. The redshift
scale shown on panel (a) was constructed assuming the WMAP
cosmological parameters from \citet{jar11}: $H_0=71.0\ \rm{km}\
\rm{s}^{-1}\ \rm{Mpc}^{-1}$, $\Omega_\Lambda=0.73$ and
$\Omega_{\rm{M}}=0.27$.

The most striking feature of the SFH shown in Panel (a) is its
complexity; it bears little resemblance to the smooth SFHs often
invoked in the theoretical models of disc evolution.  Star formation
in the Warp field began early on and proceeded at a fairly constant
rate for the first $\sim$10~Gyr, after which the rate declined
rapidly. By that point, roughly 75\% of the stars had been formed.
Between 4 and 3~Gyr ago, the SFR is consistent with being zero. The
following $\sim$1.5~Gyr saw a strong burst of star formation, about
twice the intensity of the SFR at earlier epochs, during which the
remaining 25\% of the stellar mass formed. For comparison, the
SFR per unit area of this burst at maximum is about half of the
current value in the `10~kpc ring' \citep{tab10}, the area with the
highest current SFR in M31. This burst was followed by a low
residual SFR until the present time, responsible for the sparse MS
brighter than F814W~$\sim$~26.  The derivation of the SFH using the
Padova library (see Appendix~\ref{sec:app}) is in excellent overall
agreement with that reported here.  In particular, both solutions
recover a strong burst $\sim 2-3$~Gyr ago, as well a lull in
activity $\sim 3-4$~Gyr ago, and the median age of the population,
defined as the time at which 50\% of the stellar mass was in place,
is $\sim$7.5~Gyr in both cases.

\begin{figure}          
\includegraphics[width=8.4cm]{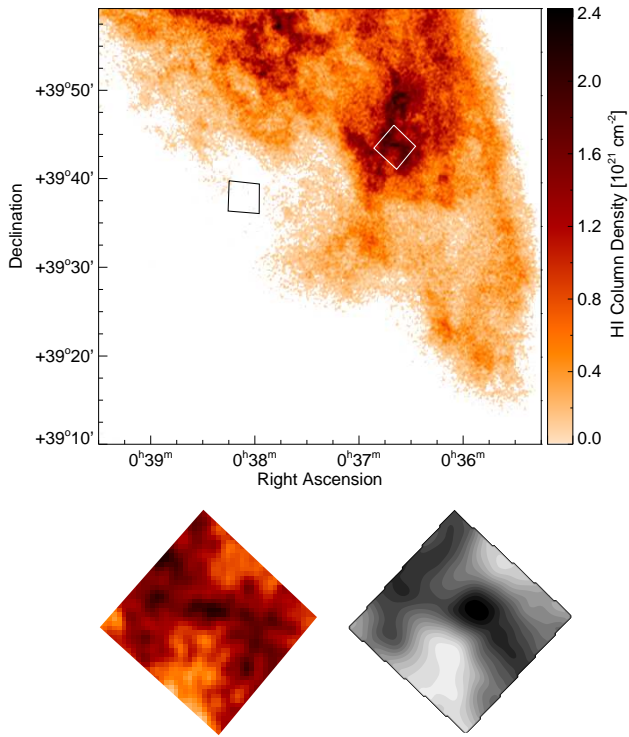}
\caption{Top: \mbox{H\,{\sc i}} column density distribution in the south-west
 disc of M31, from \citet{bra09}. The location of the Warp and Outer Disc fields
 are shown as the small black and white squares, respectively.
 Bottom: Zoom-in of the Outer Disc field. The \mbox{H\,{\sc i}} column density
 (left), and reddening map derived from the spread of the RGB stars along the
 reddening vector (right) are shown ({\it see text}). The similarity in
 morphology of the distributions is striking.
\label{fig:9}}
\end{figure}

The age-metallicity relation (AMR) shown in panel (b) is rather well
constrained, increasing smoothly from [Fe/H]~$\sim-$1.2 at 13~Gyr to
slightly above solar metallicity 2~Gyr ago.  From panel (c), the peak
metallicity [Fe/H]~$\sim-$0.7 is as expected from the morphology of
the RGB (see Figure~\ref{fig:2}), and in excellent agreement with the
stellar metallicity measured in other outer disc and halo fields both
from photometry \citep[e.g.,][]{rei04,ric04,fer05,bro06,far07,ric08}
and spectroscopy \citep[e.g.,][]{iba05,col11}.

A tantalizing feature in the AMR is the apparent decrease in
metallicity from slightly super-solar 2~Gyr ago to [Fe/H]$\sim-$0.3 at
the present-day.  This behaviour is also recovered in the solution
calculated with the Padova stellar evolution library, although not as
strong, but it does not appear in any of the tests carried out with
mock galaxies (see Appendix~\ref{sec:app}).  We explored fits in which
the metallicity was constrained to lie within a narrow range at recent
epochs but this led to larger residuals (see also \citet{bro06} who
noted a similar behaviour in their M31 fields).  The decline in
metallicity in the most recently formed populations therefore appears
genuine although it rests entirely on just a few bins.  Possible
explanations for this behaviour include a scenario in which these
stars have been accreted from a low mass galaxy \citep[e.g., a
Leo~A-type dwarf galaxy;][]{col07} or that the stars formed in situ
from low-metallicity gas which was present in this region $\sim 2$~Gyr
ago.

As shown by the Hess diagrams in panels (d)-(g), the model CMD
corresponding to this SFH matches very well the observed CMD. The only
significant residuals in the difference and significance plots are due
to the RC and red HB which, as already discussed, are notoriously
difficult features to fit with current stellar evolution models. Note,
however, that the stars of the RC and HB have not been used for the
calculations of the SFH (see Section \ref{sec:4.1}).

\subsection{Reprocessing the \citet{bar11} M33 data}

In an attempt to explain the low surface brightness substructure
recently-discovered around M33, \citet{mcc09} explored the interaction
between M31 and M33 using a set of realistic self-consistent N-body
models.  Their favoured orbits bring the systems within $\sim 50$~kpc
of each other 2--3~Gyr ago, almost exactly coincident with the major
burst of star formation we have found in the Warp field.
\citet{bar11} recently presented the SFH obtained for a deep HST/ACS
field at the edge of the main disc in M33 (R$_{deproj}$=9.1~kpc, or
about 4 scalelengths) which revealed a star formation enhancement in
the past few Gyr similar to that observed in M31.  We therefore
considered it worthwhile to further investigate the nature and
timescale of the M33 burst and examine the evidence that the
interaction could have triggered simultaneous star formation episodes.
In order to facilitate the most robust comparison, we opted to
reanalyse the M33 outer disc data (field S1 from \citet{bar11}, the
dataset for which is fully described in that paper) in the exact same
manner as the M31 data presented here. This bypasses the difficulties
inherent in the different choices of algorithms, temporal resolution,
IMF, and regions of the CMDs to be fit, as well as the representation
of the best solution itself. We thus carried out the photometry and
artificial stars tests as described in Section~\ref{sec:2} for the M33
field and calculated the SFH following the procedure laid out in
Section~\ref{sec:4.1}.

\begin{figure}          
\includegraphics[width=8.5cm]{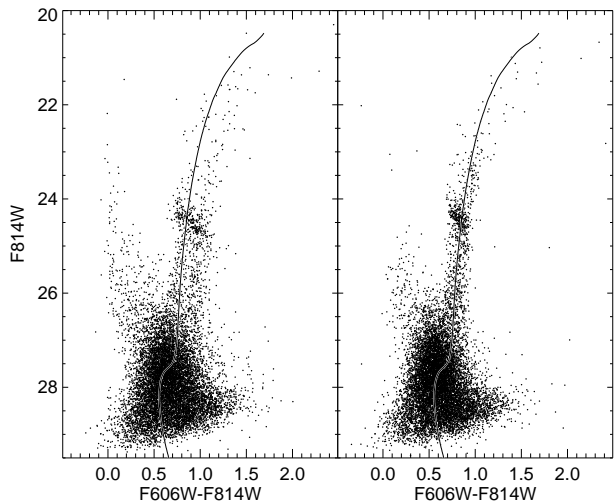}
\caption{CMDs corresponding to areas of high (left) and low (right) differential
 internal reddening in the Outer Disc field. Each CMD is drawn from $\sim 10$\%
 of the total area. The same number of stars is plotted in both
 panels. An old isochrone ([Fe/H]=$-$0.7 and 12.5~Gyr old, from the BaSTI
 stellar evolution library; \citealt{pie04}) is overplotted, assuming
 E(B-V)=0.055 from the map of \citet{sch98}.
 \label{fig:10}}
\end{figure}

The CMD is shown in Figure~\ref{fig:8c}; the reanalysis of this
dataset has led to a gain of an extra $\sim 0.5$~mag in depth over
that presented in \citet{bar11}. Despite being slightly more distant
than M31 ($\Delta (\rm{m}-\rm{M})_0\sim$~0.2, \citet{bar11}), the
somewhat longer exposure times and lower stellar density of the field
led to a similar photometric depth and 50\% completeness level as for
the Warp field.  In the left panel, the same isochrones and ZAHB as in
Figure~\ref{fig:2}, shifted to the distance and reddening of M33, have
been overplotted for comparison.  As already noted by \citet{bar11},
the oldest MSTO seen in this field is about one magnitude brighter
than the 12.5~Gyr old isochrones and therefore significantly younger
than the age of the Universe. We assume a distance of 867~kpc
\citep[i.e., (m$-$M)$_0$=24.69;][]{gal04}, inclination of 56$\degr$,
position angle of 23$\degr$ \citep{cor97}, and color excess
E(B$-$V)=0.044 as in \citet{bar11}.

In the right panel we show the bundles used to recover the SFH as
thick solid lines. These are virtually the same as the ones used for
M31 -- shown as the dotted lines, shifted to the distance and
reddening of M33. The main difference is the slight shift of the
bundle redward of the RGB to bluer colors in order to put tighter
constraints on the maximum metallicity.

Our best-fit solution for the outer disc field of M33 is shown in
Figure~\ref{fig:8b}.  For a more meaningful comparison with the SFH of
M31, panel (a) of both Figure~\ref{fig:8} and \ref{fig:8b} are
normalised to the {\it deprojected} area taking into account the
inclination and morphology of each galaxy.  The overall agreement
between our solution and that reported in \citet{bar11} is excellent:
both solutions indicate a dearth of old stars and show that less than
half of the total stellar mass was in place before redshift 0.5. The
chemical enrichment laws are also very similar, with a mild metal
enrichment from [Fe/H]~$\sim-$1.0 about 10~Gyr ago to roughly $-$0.5
at the present time, and a median [Fe/H]~$\sim-$0.7. We do find a
slightly higher SFR around $\sim$8--11~Gyr ago than that reported by
\citet{bar11} which may result from the increased CMD depth in our
reanalysis.  This finding is in better agreement with the presence of
a few RR~Lyrae stars detected in this field \citep{bar11}.  However,
the main difference between the solutions results from the finer
temporal grid which we have employed here which allows us to resolve a
strong and relatively short burst of star formation which is exactly
coeval to that in the Warp field of M31 and has the same duration
($\sim$1.5~Gyr).  We note that the drop in metallicity observed in the
SFH of the Warp field in M31 at about 2~Gyr ago is also visible in the
solution for M33.  Together, these findings reinforce the idea that
the enhanced SFRs in the outer discs of both galaxies could be related
and we discuss this possibility in more detail in Section~\ref{sec:6}.

\section{The Outer Disc Field}\label{sec:5}

As shown in Section~\ref{sec:2}, the Outer Disc field suffers from
significant differential reddening which, by blurring the features
along the reddening vector, affects the morphology of its CMD. As we
will show below, the dust distribution appears highly complex in this
region with some stars lying both in front and behind the dust clouds.
Unfortunately, this renders futile attempts to correct for the effects
of differential reddening in our data, so the method used in the
previous section to recover the SFH of the Warp field can not be
applied here.  In particular, differential reddening along the RGB
mimics the effect of higher metallicity and/or older ages by shifting
the stars to the red, thus worsening the age-metallicity degeneracy
already present.

\begin{figure}          
\includegraphics[width=8.5cm]{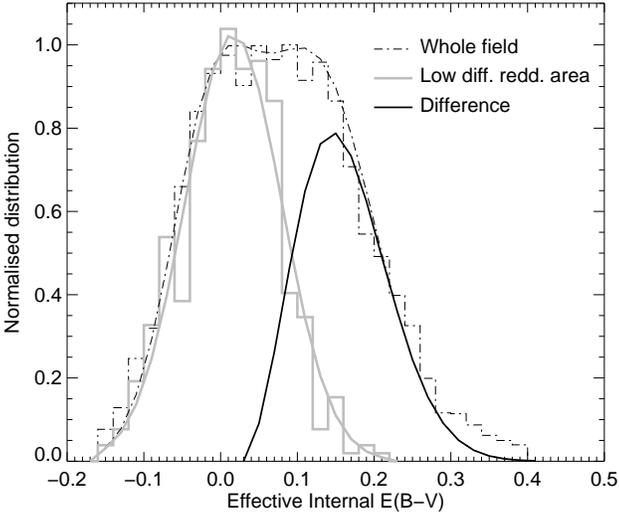}
\caption{Distribution of effective internal E(B-V) of the Outer Disc field,
 calculated from the spread of the RGB stars along the reddening vector ({\it
 see text}).  Note that this reddening is in addition to the foreground value of
 0.055 quoted earlier.
 Gaussian and double-Gaussian fits to the histograms of the low differential
 reddening area and the whole field-of-view, respectively, are overplotted. The
 thick black curve is the difference between these two Gaussian fits.
 \label{fig:11}}
\end{figure}

We focus our analysis of the Outer Disc field on quantifying the
amount of differential internal reddening, understanding its origin
and constraining the fine-scale distribution of the dust. To estimate
the amount of internal reddening as a function of position in the
field, we use RGB stars brighter than F814W~$\sim$~26.5, as shown in
the right panel of Figure~\ref{fig:3b}, since their distribution along
an old isochrone is roughly perpendicular to the reddening vector.
This allows us to keep contamination from other stellar evolutionary
phases to a minimum. We also exclude stars from the top two magnitudes
of the RGB as their color is very sensitive to small changes in
metallicity, and age to a lesser extent, as well as the part of the
RGB which overlaps the RC.

Given that the blue side of the RGB is well represented by a 12.5~Gyr
old isochrone with [Fe/H]=$-$0.7 (see Figure~\ref{fig:2}), we assume
that all RGB stars should have been observed at this location if
differential reddening had been negligible. Under this assumption, we
can estimate the amount of extinction along the line of sight to a
given RGB star by measuring the distance, along the reddening vector,
between the star and this isochrone, and convert this distance to an
`effective internal E(B-V)'. Note that this represents the {\it internal}
reddening as it is in addition to the foreground value of
E(B$-$V)$=0.055$ reported in Table \ref{tab1} and taken from the map
of \citet{sch98}.

In the bottom right panel of Figure~\ref{fig:9} we show the resulting
contour map, where darker tones indicate higher extinction. It shows
that the distribution of reddening is far from homogeneous in this
small field-of-view (FOV) and delineates clumps and filaments with
sizes $\la$ a few hundred parsecs.  Interestingly, the distribution of
reddening is almost an exact match to the morphology of the
\mbox{H\,{\sc i}} column density map \citep{bra09}, kindly supplied by
R. Braun and shown in the top and bottom left panels of
Figure~\ref{fig:9}. By comparing the low resolution \mbox{H\,{\sc i}}
observations of \citet{new77} with their photometry, \citet{cui01}
also found that the colors of the RGB stars were correlated with the
column density of \mbox{H\,{\sc i}} in their
28$\arcmin\times$28$\arcmin$ outer disc field -- which includes our
HST field.  They concluded that the relatively large amount of
extinction is closely associated with the \mbox{H\,{\sc i}} gas, and
that the outer disc of M31 therefore contains substantial amounts of
dust.  Our result shows that this interpretation is still valid on the
much finer scales sampled here.

From our contour map, we can now select the areas within the FOV where
extinction is lowest and highest. We pick stars which fall within two
rectangular boxes, each covering roughly 10\% of the total area of the
ACS FOV.  Figure~\ref{fig:10} shows the CMDs of these two fields, each
containing the same number of stars. The isochrone described above is
overplotted in each panel, shifted to the distance of M31 and
corrected for foreground reddening. In the left panel, the RGB is very
wide, almost bimodal, with a blue RGB nicely fit by the isochrone and
a second, redder RGB. The presence of both non- and very reddened
stars at the same spatial location indicates a complex 3-dimensional
spatial distribution where non-reddened stars possibly lie in front of
the dust cloud and the very reddened stars behind.  This is precisely
why we cannot correct the photometry for differential reddening:
applying a correction derived from the contour map of
Figure~\ref{fig:9} would simply shift the wide RGB to the blue, rather
than producing a narrow RGB.

In the right panel of Figure~\ref{fig:10}, the isochrone provides
a good fit to the stars of the lowest differential reddening area,
which shows that the differential reddening in this subfield is more
limited.  For simplicity, we assume in the following that this CMD
is representative of the intrinsic spread in color due to a
combination of photometric errors and ranges of age and metallicity.
However, we note that the lower RGB here is still not as tight as in the
Warp field, and thus some differential reddening is still likely to
be present if we assume the intermediate-age and old stellar
populations to be comparable between the Warp and Outer Disc
fields.

Figure~\ref{fig:11} shows the normalised distribution of `effective
internal E(B$-$V)', as calculated above, for the low differential
reddening area (gray histogram) and the whole Outer Disc field
(dash-dotted black histogram). Gaussian and double-Gaussian fits to
these histograms, respectively, are also shown.  The difference
between the two therefore represents the additional contribution of
the differential reddening to the spread in the CMD, and is shown as
the thick black curve in Figure~\ref{fig:11}.  Note that the
distribution of E(B$-$V) of the low differential reddening area in
Figure~\ref{fig:11} is not centred on zero, but is slightly offset by
$\sim$0.02.  While this might indicate that the foreground reddening
taken from the \citet{sch98} map at this location is underestimated,
it is worth stressing that the values of the `effective internal
E(B$-$V)' are only relative to a given isochrone which may not be a
perfect representation of the true local reddening. Our goal here is
to quantify the differential internal reddening rather than that due
to the foreground.  The distribution under the thick black curve has a
median of $\sim$0.16; subtracting the 0.02 offset, we find that this
additional extinction amounts to $\sim$0.14 (corresponding to a total
reddening of $\sim0.2$) and affects about 40\% of all the RGB
stars\footnote{Note that there is no feature in the Schlegel maps at
the location of the Outer Disc field. This could be due to the low
effective resolution of these maps which averages over structures,
or the fact that outer disc dust is too cold to emit significantly
in the DIRBE bandpasses.}.

While we cannot do a detailed SFH reconstruction for the Outer Disc
field, we can make an important deduction about the recent SFH from
inspection of the high and low reddening CMDs in Figure~\ref{fig:10}.
Neither of these CMDs show evidence for the prominent 1--2 Gyr
turn-off population so readily apparent in the Warp CMD hence this
region apparently did not undergo a recent burst.  Furthermore, the
youngest populations ($\sim$ few tens of Myr) are largely confined to
the highest reddening regions of the FOV.  That such stars are present
in the Outer Disc field and not at all in the Warp is not surprising
when considering the gas distribution (see Figure \ref{fig:9}) which
shows that the stellar warp region is currently devoid of high-column
density gas.

\section{Variable Stars}\label{sec:3}

The candidate variable stars were extracted from the photometric
catalogs using the \citet{ste96} variability index $J$, which uses the
correlation in brightness change in paired frames to isolate possible
variables from constant stars. This process resulted in the selection
of $\sim $ 180 and 230 candidates for the Warp and Outer Disc fields,
respectively. Most of these turned-out to be false detections, owing
their non-zero variability index to the combination of sparse sampling
and large numbers of cosmic rays. Inspection of the individual light
curves, and the location of the candidates on the stacked images and
on the CMDs, allowed us to extract the candidate RR~Lyrae stars. The
other bona fide variables, mostly eclipsing binary systems, were
discarded due to the sparse sampling preventing us from finding their
periods. Our final catalog of variable stars contains 12 and 13
RR~Lyrae stars from the Warp and Outer Disc fields, respectively. All
of these are new discoveries, since no data of sufficient depth or
sampling has been obtained before our observations of this part of
M31. Interestingly, no Cepheids were detected, despite the presence of
young, massive stars in the Outer Disc field.

We performed the period search through Fourier analysis \citep{hor86},
which was then refined by hand to obtain smoother light curves using
the same interactive program as described in \citet{ber09}. However,
the small number of datapoints, as well as an inconvenient temporal
sampling (all the images in one band observed consecutively, followed
by all the images in the other band) significantly complicated the
process.  For some variables, several periods could fold our data and
give light curves of similar quality. We thus used additional
parameters to constrain the most likely period: an amplitude ratio
A$_{F606W}$/A$_{F814W}\sim$~1.8, and a reasonable location on both the
period-amplitude diagram and the CMDs. These constraints were
determined from the other RR~Lyrae stars for which the period search
was straightforward.  While this may not be ideal for obtaining purely
objective periods, the process allowed us to obtain reasonably
reliable estimates given the characteristics of the data. It is
important to note that some of our final periods could well be aliases
of the true period.

\begin{table*}
 \centering
 \begin{minipage}{130mm}
  \caption{Properties of the Variable Stars in the Warp field\label{tab2}}
  \begin{tabular}{@{}ccccccccc@{}}
  \hline
ID & Type & R.A. & Decl. & Period & $\langle F606W \rangle$ & A$_{F606W}$ &
$\langle F814W \rangle$ & A$_{F814W}$\\
 & & (J2000) & (J2000) & (days) & & & &\\
 \hline
  V01 & $c$  & 00 38 02.64 & +39 36 30.9 & 0.46  & 26.03 & 0.35 & 25.63 & 0.39 \\
  V02 & $ab$ & 00 38 10.16 & +39 36 42.8 & 0.49  & 25.40 & 1.25 & 25.03 & 0.67 \\
  V03 & $ab$ & 00 38 03.65 & +39 37 23.9 & 0.47  & 24.86 & 0.65 & 24.46 & 0.63 \\
  V04 & $ab$ & 00 38 13.85 & +39 37 37.0 & 0.60  & 25.79 & 0.71 & 25.39 & 0.29 \\
  V05 & $ab$ & 00 38 13.53 & +39 38 03.9 & 0.63  & 25.50 & 0.69 & 24.94 & 0.43 \\
  V06 & $ab$ & 00 38 09.02 & +39 38 13.8 & 0.70  & 25.03 & 0.81 & 24.52 & 0.37 \\
  V07 & $ab$ & 00 38 03.12 & +39 38 20.3 & 0.50  & 25.12 & 1.44 & 24.75 & 0.71 \\
  V08 & $ab$ & 00 38 03.05 & +39 39 07.5 & 0.58  & 25.00 & 0.79 & 24.56 & 0.46 \\
  V09 & $ab$ & 00 38 13.77 & +39 39 08.2 & 0.53  & 24.93 & 1.18 & 24.44 & 0.75 \\
  V10 & $ab$ & 00 38 11.26 & +39 39 13.3 & 0.78  & 24.77 & 0.41 & 24.31 & 0.22 \\
  V11 & $c$  & 00 38 11.11 & +39 39 14.2 & 0.305 & 25.68 & 0.27 & 25.12 & 0.32 \\
  V12 & $c$  & 00 38 13.76 & +39 39 16.8 & 0.378 & 24.92 & 0.58 & 24.58 & 0.27 \\
\hline
\end{tabular}
\end{minipage}
\end{table*}

\begin{table*}
 \centering
 \begin{minipage}{130mm}
  \caption{Properties of the Variable Stars in the Outer Disc field\label{tab3}}
  \begin{tabular}{@{}ccccccccc@{}}
  \hline
ID & Type & R.A. & Decl. & Period & $\langle F606W \rangle$ & A$_{F606W}$ &
$\langle F814W \rangle$ & A$_{F814W}$\\
 & & (J2000) & (J2000) & (days) & & & &\\
 \hline
  V13 & $ab$ & 00 36 35.63 & +39 42 17.0 & 0.56  & 25.88 & 1.01 & 25.15 & 0.94 \\
  V14 & $c$  & 00 36 34.10 & +39 42 27.9 & 0.269 & 26.34 & 0.46 & 25.76 & 0.18 \\
  V15 & $c$  & 00 36 34.29 & +39 43 02.9 & 0.35  & 26.25 & 0.32 & 25.73 & 0.33 \\
  V16 & $ab$ & 00 36 43.61 & +39 43 10.0 & 0.52  & 25.07 & 1.18 & 24.66 & 0.77 \\
  V17 & $c$  & 00 36 42.00 & +39 43 20.4 & 0.343 & 25.06 & 0.40 & 24.71 & 0.24 \\
  V18 & $ab$ & 00 36 28.25 & +39 43 29.9 & 0.59  & 24.72 & 1.03 & 24.36 & 0.36 \\
  V19 & $c$  & 00 36 36.18 & +39 43 40.6 & 0.35  & 25.85 & 0.69 & 25.30 & 0.44 \\
  V20 & $c$  & 00 36 34.25 & +39 43 40.9 & 0.405 & 25.91 & 0.21 & 25.12 & 0.23 \\
  V21 & $c$  & 00 36 30.60 & +39 43 58.5 & 0.38  & 25.69 & 0.51 & 25.12 & 0.43 \\
  V22 & $ab$ & 00 36 41.28 & +39 44 01.9 & 0.54  & 25.39 & 0.55 & 25.01 & 0.39 \\
  V23 & $c$  & 00 36 44.44 & +39 44 41.4 & 0.39  & 24.97 & 0.60 & 24.62 & 0.35 \\
  V24 & $ab$ & 00 36 42.72 & +39 45 10.2 & 0.452 & 25.15 & 1.17 & 24.88 & 0.77 \\
  V25 & $ab$ & 00 36 40.05 & +39 45 48.8 & 0.51  & 24.92 & 1.01 & 24.54 & 0.68 \\
\hline
\end{tabular}
\end{minipage}
\end{table*}

\begin{figure}          
\includegraphics[width=9cm]{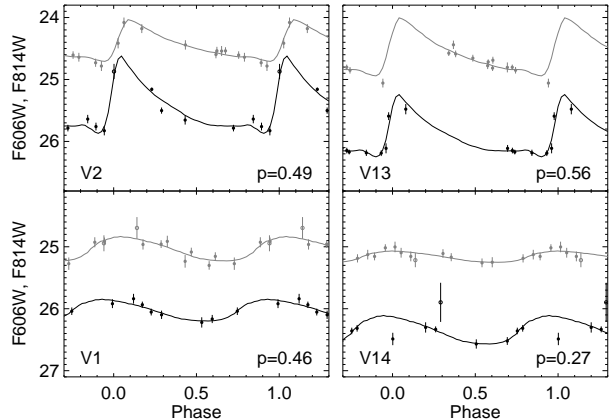}
\caption{Sample light-curves of two RR$ab$ ({\it top}) and two RR$c$
 ({\it bottom}), taken from the Warp ({\it left}) and Outer Disc
 ({\it right}) fields, illustrating the sparse sampling of our data.
 Black and grey correspond to the F606W and F814W magnitudes,
 respectively. The star ID is indicated in the lower left and the
 period in days in the lower right.}
\label{fig:4}
\end{figure}

Sample light curves for fundamental and first-overtone mode pulsators
from each field are shown in Figure~\ref{fig:4}, where templates from
the set of \citet{lay99} have been overplotted. Given the difficulty
of measuring accurate periods and mean magnitudes, as well as the
incompleteness due to the limitations outlined above, we do not
attempt to analyse the detailed individual and global properties of
these stars. Their locations and approximate properties are given here
for future reference.

The properties of the RR~Lyrae stars are summarised in
Table~\ref{tab2} and 3 for the Warp and Outer Disc fields,
respectively. The first and second columns give the identification
number and the type of pulsation, with $ab$ and $c$ designating the
fundamental (RR$ab$) and first-overtone (RR$c$) mode pulsators,
respectively. The discrimination between types was based on the period
and light curve morphology. The next two columns contain the
equatorial coordinates in J2000.0, while columns (5) lists the
period in days. Except in the
few cases where a third digit resulted in a significantly smoother
light curve (mostly RR$c$), we only give the periods with two
significant figures.  In the remaining columns, the intensity-averaged
mean magnitude, as well as the amplitude of the pulsations,
are listed.

The detection of RR~Lyrae variables this far out in the disc shows how
the variability studies can complement the deep CMD analyses: none of
the CMDs presented in this paper harbor the obvious features that are
traditionally associated with truly old stellar populations, such as a
well-defined faint MSTO or an extended HB. Yet, the characteristic
properties of these variables demonstrate that at least some of the
stars at F606W-F814W$\sim$0.5 and F814W$\sim$25 belong to an old,
extended HB.

\section{Discussion}\label{sec:6}

The CMD-fitting technique applied to the Warp field reveals two main
episodes of star formation: an extended period of roughly constant
star formation over the period from 4.5 to 13~Gyr ago (corresponding
to z$\ga 0.5$), and a relatively short-lived burst which peaked
$\sim2$~Gyr ago.  In the $\sim$1~Gyr which separates these episodes,
very little star formation took place.

The differential reddening present in the Outer Disc field
prevents a similar analysis of its SFH.  Although we cannot deduce
anything about the SFH at early times in this field, the lack of a
prominent 1--2 Gyr old MSTO feature in the CMD indicates that it did
not experience the same recent burst as seen in the Warp.  It is
interesting to speculate why the two fields could differ in this
respect.  Firstly, the disc of M31 is strongly perturbed at large
radius and highly inclined with respect to the line of sight.  Thus,
even though the two pointings appear close in projection on the sky
they may actually sample very different radii.  If the SFH over the
last few Gyr varies strongly as a function of radius in the M31
outer disc, this could explain the different behaviours observed.
Furthermore, if the two fields sample different disc radii then they
will also have differing levels of contamination from other
structural components (e.g. thick disc, stellar halo) which could
impact the overall SFHs.  There is some evidence that this might be
the case from examining published kinematical data for RGB stars in
the M31 outer disc.  Although not exactly coincident with our ACS
fields, fields F1 and F2 from \citet{iba05} are in good proximity to
the Outer Disc and Warp pointings respectively.  While field F1
shows a single velocity peak with a velocity lag and moderate
dispersion ($\sim30$ km/s), field F2 shows the same component plus a
dominant cold component ($\sim 9$km/s) which exactly matches the
expected circular velocity of the disc at that location.  This
suggests that the thin disc may be more dominant in the Warp field
than in the Outer Disc field, with the thick disc dominating in the
latter.

In what follows, we discuss the two main episodes of star formation in
the Warp field separately, assuming that the behaviour observed here
is representative of the whole outer thin disc.

\subsection{Star Formation at Intermediate and Early Epochs}

The star formation rate at lookback times of $\ga 4.5$~Gyr ago (z$\ga
0.5$) is roughly constant and does not show salient features (see
Figure~\ref{fig:8}). While this behaviour may be genuine, it could
also (partly) result from the lack of resolution at faint magnitudes
(hence old ages) which tends to smooth the resulting SFH (see the
tests on mock galaxies in the Appendix as well as the discussion in
\citealt{hid11}).  One consequence of this smoothing is that some star
formation is always recovered in the oldest age bins, even although
none may have taken place.  Nonetheless, the presence of RR Lyrae
stars in the Warp field provides support for a population of stars
with age $\ga 10$~Gyr.

If the stars in the Warp field formed in situ, the significant median
age of 7.5~Gyr at $\sim 5$ radial scalelengths indicates that M31's
outer disc was in place early on.  This argues against a scenario in
which M31 underwent a major merger at intermediate epochs after which
the disc reformed, as has been recently suggested by \citet{ham10}.
Other evidence also supports the idea that the outer discs of massive
spiral galaxies formed at high redshift
\citep[e.g.,][]{fer01,bro06,wil09a,mun11}.  On the other hand, the
outer disc of M33 lacks a significant fraction of old stars at 3--4
scalelengths \citep{wil09b,bar11}.  This could reflect the effect of
downsizing within the disc population, whereby stars in more massive
discs tend to have formed earlier and over a shorter time-span than in
less massive systems \citep[e.g.,][]{cow96}.  Alternatively, it may
simply highlight the uniqueness of M33 -- indeed \citet{gog10} find
that the outer disc of the similarily low-mass spiral galaxy NGC~300
is dominated by old stars.

Another interesting feature is the fact that the dominant population
which formed at early epochs is relatively metal-rich
([Fe/H]$\ga-$1.3, Figure~\ref{fig:8}c), at least in the solution
obtained with the BaSTI stellar evolution library.  This is
qualitatively similar to the results of \citet{wil09a} for the outer
disc of M81, which is to date the only other massive disc galaxy for
which the CMD-fitting technique has been applied (although their data
do not reach old main-sequence turn-offs). The solutions for their
field, located at five scalelengths from the centre of M81, show a
constant [M/H]~$\sim-$0.5 at all ages. Similarly, our renalysis of the
\citet{bar11} M33 outer disc field supports the lack of stars with
[Fe/H]~$\la-$1.5, consistent with their original findings. Therefore,
it seems that even the oldest stars in the outer discs of galaxies
must have formed from pre-enriched gas.

In the discussion above, we have assumed that the stars which we see
today in the outer disc of M31 (and M33) actually formed there.  It is
possible, however, that these stars formed at much smaller
galactocentric distances, where the metallicity was higher at early
epochs due to the higher rate of gas recycling, and migrated to their
current location due to resonant scattering with transient spiral arms
and bars \citep[e.g.,][]{sel02,ros08a,ros08b,sch09,min11} or other
mechanisms \citep{qui09,bir11}.  For example, in the models of
\citet{ros08b}, roughly 50\% of solar neigbourhood stars (i.e. those
lying at $\sim 2-3$ scalelengths) have come from elsewhere, with the
bulk originating at smaller radii and moving outwards.  At extreme
radial locations such as we have probed here ($4-6$ scalelengths),
this fraction of migrated stars is likely to be even higher on
account of the low rate of {\it in situ} star formation.  Indeed,
both M31 and M33 exhibit spiral arm patterns which extend to these
large radii, indicating that migration could still be an efficient
mechanism for moving stars around.  The net effect of radial
migration on outer disc stellar populations is to render the SFH and
AMR flat and featureless in these parts (see Figure 3 of
\citet{ros08b}). This behaviour is in stark contrast to what we see
in the Warp and M33 S1 fields, where the SFHs display sharp
features, at least back to $\ga$5~Gyr ago, and the AMRs increase
smoothly with time from early epochs to 2~Gyr ago.  This suggests
only a small role for radial migration in populating the M31 and M33
outer discs at the radii we have probed.  For the same reasons, it
is also unlikely that a significant fraction of the stars in these
parts have been directly accreted onto the disc plane from satellite
galaxies.

\subsection{The 2~Gyr Old Burst}

Strong bursts of star formation are often caused by an external
trigger such as an accretion event or an interaction with a nearby
galaxy \citep[e.g.,][]{ken87}.  Indeed, the copious amount of
substructure observed in the outer regions of M31 strongly suggests
interactions have taken place \citep[e.g.,][]{fer02,ric08,mcc09}.

The well-known giant stream in the south of M31 \citep{iba01} is the
fossil evidence of one such event. It has been modeled rather
successfully as either the remnant of a $\sim$$10^9M_{\sun}$ satellite
accreted less than a billion years ago \citep[e.g.,][]{fard07}, or a
major merger in which the interaction and fusion with M31 may have
started $\sim 9$ and 5.5~Gyr ago, respectively \citep{ham10}.  Neither
of these scenarios can account for a star formation burst that occured
about 2~Gyr ago in the Warp field.

Another possibility is that the close passage of M33 could have
triggered star formation on large scales in M31. Using realistic
self-consistent N-body simulations, \citet{mcc09} were able to best
reproduce the morphology of the outer substructure in M33 with orbits
that brought the systems within $\sim$50~kpc of each other 2--3~Gyr
ago and that satisfy their current positions, distances, radial
velocities, and M33's proper motion.  In particular, they found that
M33 reached pericenter about 2.6~Gyr ago, which is in excellent
agreement with the start of the star formation bursts observed in
Figures~\ref{fig:8} and \ref{fig:8b}. This is highly suggestive of an
interaction between the two systems triggering inward flows of
metal-poor gas which fuelled moderate bursts of star formation. Such
behaviour is consistent with the predictions of state-of-the-art
simulations of galaxy interactions and mergers, even though most work
to date has focused on the inflow of gas to the central regions
\citep[e.g., \citealt{dim08},][]{tey10}.

Further development of these N-body models (Dubinski et al.\, in
preparation) indicates that the gravitational interaction of M33
should have induced perturbations in M31 in the form of heating and
disruption of the outer disc.  This is consistent with the presence of
warps in both the \mbox{H\,{\sc i}} and stellar discs, and the
disc-like stellar populations found in many regions of substructure
\citep{fer05,far07,ric08}. Such an interaction could also help explain
the large fraction of globular-like clusters younger than 2~Gyr in M31
\citep{fan10}, the enhanced SFR 2--4~Gyr ago observed in all the deep
{\it HST} fields in M33 \citep{bar07,wil09b,bar11}, and the warps in
both the \mbox{H\,{\sc i}} \citep[e.g.,][]{wri72,put09} and stellar
discs \citep{mcc10} of M33.  To put this interpretation on firmer
footing, additional data are required to assess just how widespread
the $\sim$2~Gyr burst of star formation is in the M31 outer disc and
address how the timing and intensity of the burst varies with
location.

\section{Conclusions}\label{sec:7}

We have presented the analysis of very deep, multi-epoch {\it HST}
data for two fields located in the far outer disc of M31 at $\sim
2\degr$ (26~kpc) from the centre, corresponding to 5--6 radial
scalelengths.

We apply the CMD-fitting technique to the Warp field, and find that a
significant fraction of the stellar mass ($\sim$25\%) was produced as
the result of a strong and relatively short-lived burst of star
formation about 2~Gyr ago. The re-analysis of the M33 outer disc S1
field of \citet{bar11} and resulting SFH show a similar burst of SF
that is exactly coeval with that observed in M31. The burst in both
galaxies is also accompanied by a decline in global metallicity, which
suggests an inflow of metal-poor gas into the outer disc.  These
results further support the inference from N-body modelling that M31
and M33 had a close interaction about 2--3~Gyr ago that enhanced the
outer disc SF at that epoch, and may also be responsible for the warps
in the \mbox{H\,{\sc i}} and stellar discs of both galaxies and the
large fraction of clusters younger than about 2~Gyr old.

Further back, the SFH was roughly constant over the period spanning
4.5--13~Gyr. The presence of stars older than about 10~Gyr, while
surprising at such a large distance from the nucleus, is confirmed by
the detection of RR~Lyrae stars in this field. In addition, the sharp
features in the SFH and the rather well-defined and monotonically
increasing AMR favour the {\it in situ} formation of these stars,
rather than an outward migration from birthplaces at much smaller
galactocentric radii.  The fact that the oldest populations are still
fairly metal-rich ([Fe/H]$\ga-$1.3) suggests pre-enrichment
of the outer disc material.

The second field, the Outer Disc field, is strongly affected by
differential reddening which prevents us from applying the CMD-fitting
technique. However, the pattern of reddening inferred from the color
of the RGB stars is highly correlated with the \mbox{H\,{\sc i}}
distribution in the outer disc, suggesting these clouds contain
substantial dust indicative of significant metal enrichment.  This
confirms earlier work by \citet{cui01} but goes further in
demonstrating how highly structured the dust distribution is on 
small scales.  The presence of significant reddening in these parts is
consistent with the roughly solar metallicity in the past few Gyr
uncovered by our SFH analysis of the nearby Warp field.

Finally, from a time-series analysis, we find a dozen RR~Lyrae stars
in each of the two fields.  The detection of these variables supports
the presence of truly old stellar populations in the far outer discs
of galaxies despite the lack of obvious HBs in the CMDs, and shows how
variability studies can complement deep SFH analyses.

 \section*{Acknowledgments}

 Support for this work was provided by a Marie Curie Excellence Grant
 from the European Commission under contract MCEXT-CT-2005-025869,
 a rolling grant from the Science and Technology Facilities Council,
 and the Ministry of Science and Innovation of the Kingdom of Spain
 (grant AYA2010-16717).
 The authors are grateful to Jenny Richardson, Maud Galametz and
 Victor Debattista for interesting discussions and Robert Braun for
 providing the HI data and useful comments. We would like to thank
 the referee for a detailed report that helped improve
 this manuscript. AMNF acknowledges support from a Caroline Herschel
 Distinguished Visitorship and the Institute of Astronomy, Cambridge.
 SLH and MM are supported by the IAC (grant 310394), and the Science
 and Innovation Ministry of Spain (grants AYA2007-3E3507 and
 AYA2010-16717). GFL thanks the Australian Research Council
 for support through his Future Fellowship (FT100100268) and Discovery
 Project (DP110100678).

\appendix

\section{Testing the SFH Calculation Method}\label{sec:app}

Here we describe a series of tests aimed at understanding how reliable
the CMD-fitting technique used in this paper is, given the photometric
properties of our dataset (i.e., depth, crowding), as well as
revealing possible biases and limitations of the method.

\begin{figure*}          
\begin{center}
\includegraphics[width=15.cm]{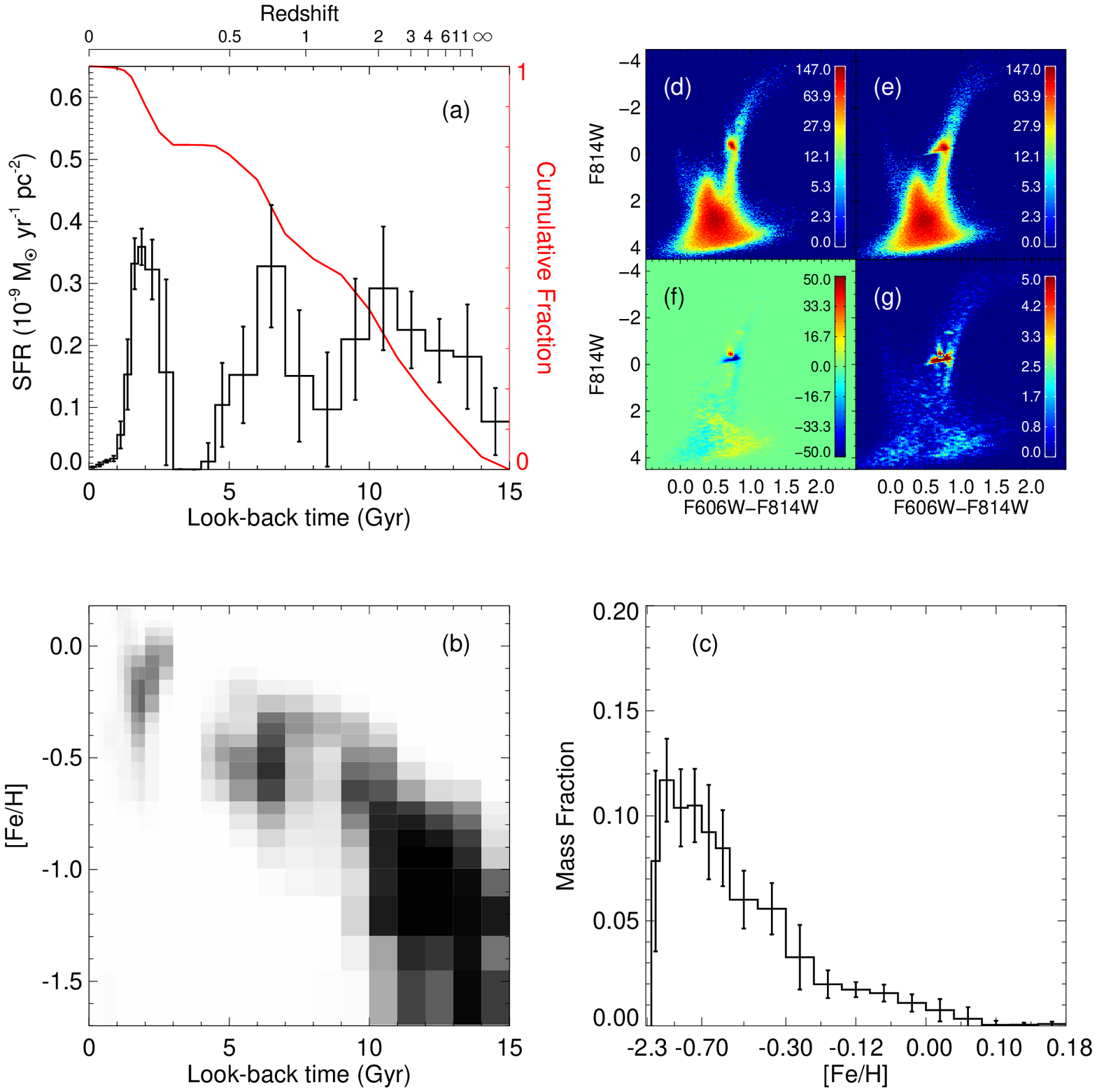}
\caption{Same as Figure~\ref{fig:8}, but recovered with the Padova stellar
 evolution library.
\label{fig:a1}}
\end{center}
\end{figure*}

The first test consisted of recovering the SFH of the Warp field in
the exact same way as described in Section~\ref{sec:4}, albeit using
the stellar evolution library of the Padova group \citep{gir00}.  The
resulting SFH is shown in Figure~\ref{fig:a1}. The agreement with the
SFH obtained with BaSTI (see Figure~\ref{fig:8}) is excellent. The
main features such as the burst centered on 2~Gyr old, the very low
star formation between 3--4~Gyr and the smooth increase of metallicity
from the earliest epochs to 2~Gyr ago are virtually the same as with
the BaSTI library, and the recent decrease in metallicity is also
observed, albeit not as strongly. The main difference lies in the
overall metallicity, which appears systematically lower by roughly
0.2~dex in the former solution at any age. The sense and amount of
this offset is exactly the same as reported in \citet{bar11}, and is
likely due to the difference in color of the RGB between the two
libraries \citep[e.g.,][]{gal05}. We note that the SFH obtained by
\citet{bro06} with the Victoria--Regina isochrones \citep{van06} for
another outer disc field in M31 shows very few stars with metallicity
lower than about [Fe/H]=$-$1, which is more in line with the SFH we
obtained with BaSTI and agree with spectroscopic estimates of outer
disc metallicities in the literature \citep{iba05,col11}.  Another
difference between the solutions is that the star formation which
occurred earlier than 5~Gyr is not as uniform in the Padova solution
as in the BaSTI one, but shows a mild decrease about 8--9~Gyr ago.
Given the larger errorbars in the solution obtained with the Padova
library, however, we do not consider this feature to be very
significant.

\begin{figure*}          
\begin{center}
\includegraphics[width=15.cm]{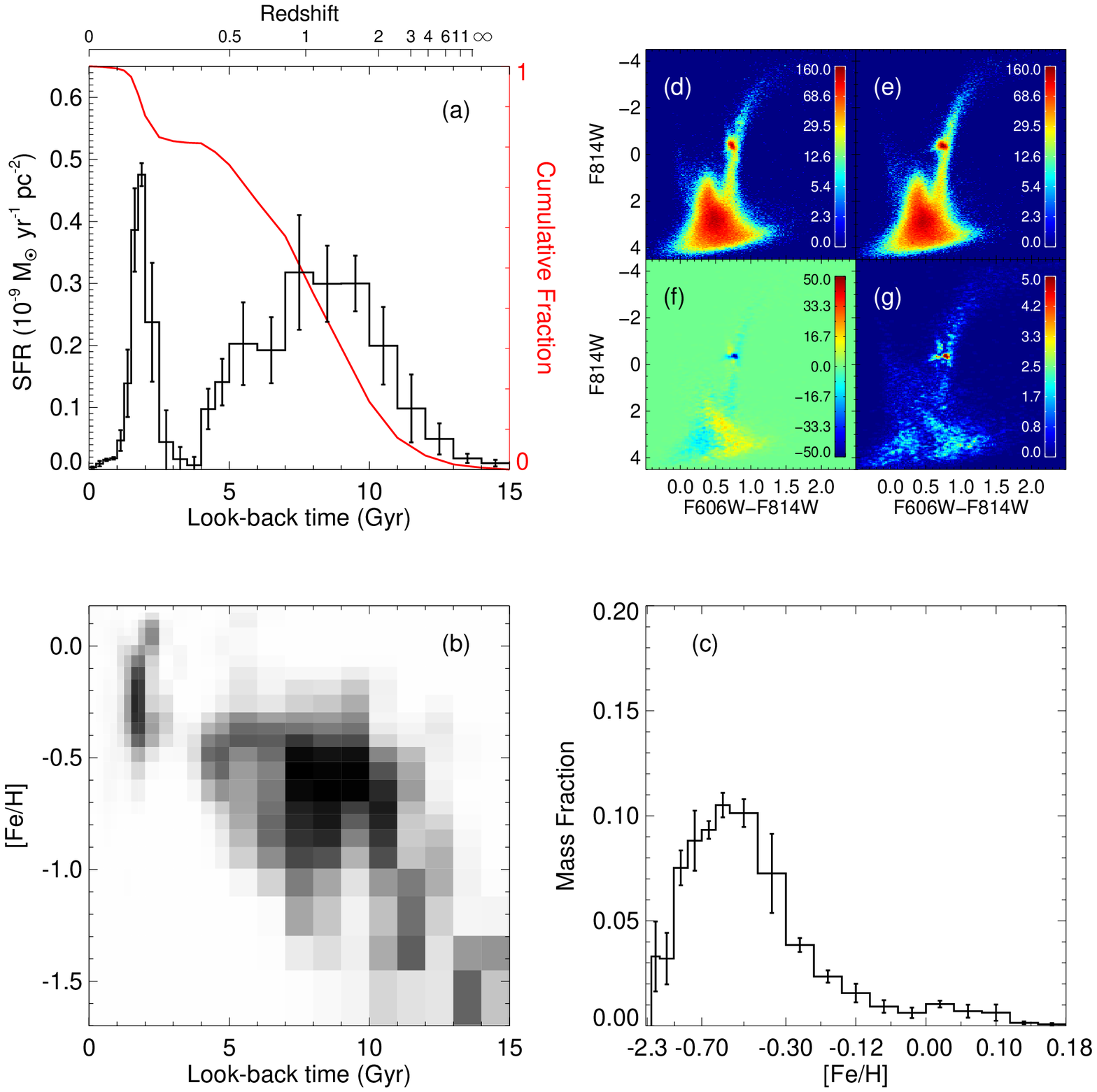}
\caption{Same as Figure~\ref{fig:8}, but using two additional bundles to
 include the RGB, RC, and HB in the fitting, as shown in Figure~\ref{fig:3b}.
\label{fig:a8}}
\end{center}
\end{figure*}

For the second test we repeat the fit of the Warp CMD with the BaSTI
library, but this time including the evolved stars -- as shown by
the dashed bundles in Figure~\ref{fig:3b}.
The solution presented in Figure~\ref{fig:a8} is {\it
qualitatively} similar but the fit is significantly worse
($\overline{\chi^2}$=5.27). This is evidenced by the significant
residuals in the main-sequence shown in panels (f) and (g).  We
believe the simultaneous fit of all the features of the CMD
produces worse results due to the larger uncertainties in the
stellar evolution models of the evolved stars.

\begin{figure*}          
\begin{center}
\includegraphics[width=15.cm]{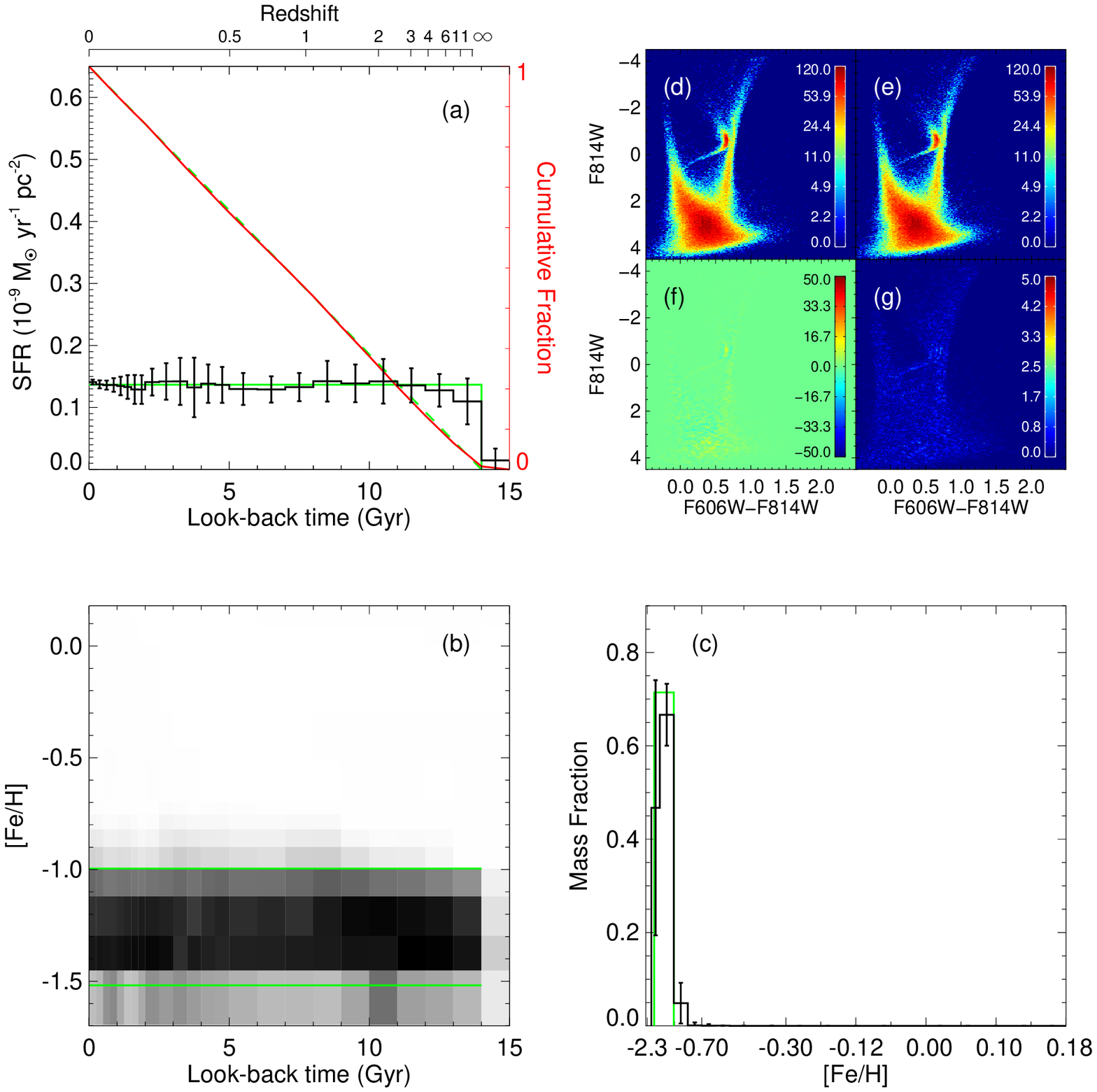}
\caption{Mock galaxy with constant SFR, created and recovered with the BaSTI
 stellar evolution library. The green solid lines in panels (a) and (c) show the
 input SFH, while in panel (b) they delineate the upper and lower limits of the
 input metallicity range, between which the stars were uniformly distributed in
 each age bin. The green dashed line in panel (a) is the cumulative fraction of
 the input SFH.
\label{fig:a2}}
\end{center}
\end{figure*}

The remaining tests involve `mock galaxies' for which CMDs were
produced using IAC-star \citep{apa04} and invoked SFHs representative
of the typical range seen within the Local Group.  Observational
effects were simulated using the results of the artificial stars tests
described in Section~\ref{sec:2.4}.  Figure~\ref{fig:a2} shows the
simplest possible SFH, that of a constant rate between 14~Gyr ago and
the present time without chemical enrichment.
In panel (a), the input SFH is represented by the green lines, while
the recovered SFH is shown as in the previous figure. In panel (b) the
two green lines show the upper and lower limits of the input
metallicity range, between which the stars were uniformly distributed
in each age bin.
Ideally, discrepancies between the input and recovered SFH should
be representative of the effects of the observational
effects and age-metallicity degeneracy.

\begin{figure*}          
\begin{center}
\includegraphics[width=15.cm]{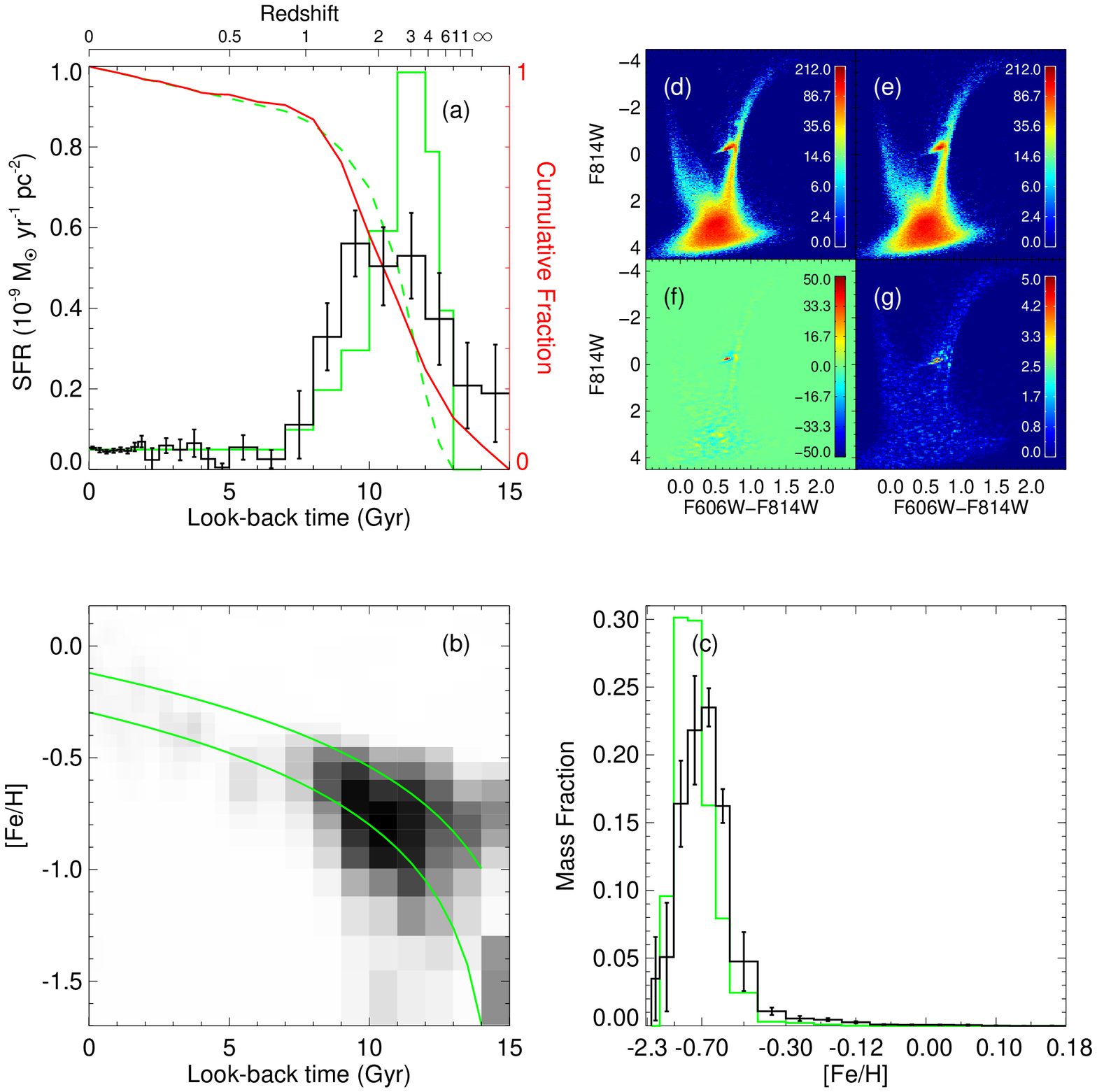}
\caption{Same as Figure~\ref{fig:a2}, but with a dSph-like SFH.
\label{fig:a4}}
\end{center}
\end{figure*}

\begin{figure*}          
\begin{center}
\includegraphics[width=15.cm]{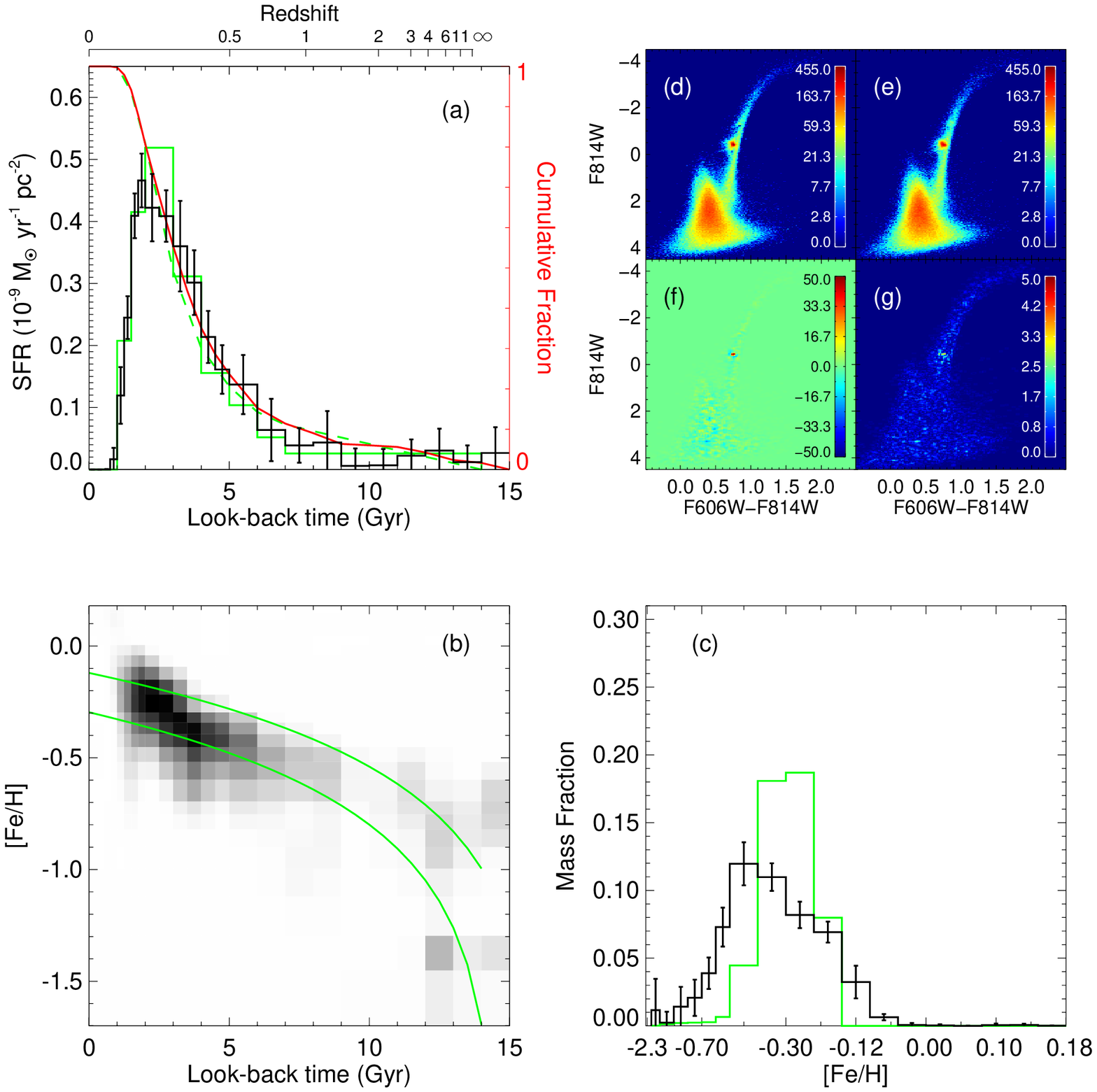}
\caption{Same as Figure~\ref{fig:a2}, but with a Leo~A-like SFH.
\label{lastpage}
\label{fig:a5}}
\end{center}
\end{figure*}

Finally, for the remaining two mock galaxies, we test the capacity to
recover the SFH of a predominantly old galaxy similar to the
transitional dIrr/dSph of the Local Group \citep[e.g.,
LGS3:][]{hid11}, and of a mostly young and intermediate-age galaxy
like the dwarf irregular Leo~A \citep{col07}, both with a mild
chemical enrichment over the past 14~Gyr.
The results are shown in Figure~\ref{fig:a4} and \ref{fig:a5}.

Overall, the SFHs are recovered very well in all cases. Not only are
the residuals in the CMDs very low, but the SFRs and metallicities are
very close to the input values. In addition, the AMR is measured
correctly, which highlights the ability of the method to solve for
both age and metallicity simultaneously. There are small discrepancies
however, which appear to be systematic:

\begin{itemize}
\item while the correct mean metallicity at a given age is usually
  recovered, the dispersion of metallicity is often
  overestimated, in the sense that stars are recovered outside the
  range delineated by the green lines in panel (b) of
  Figures~\ref{fig:a2} to \ref{fig:a5};
\item this effect is strongest for the most recent SFH ($\la$1~Gyr)
  since metallicity only has a small effect on the color of the bright
  main sequence and is therefore largely unconstrained;
\item the ages recovered for the oldest populations ($\ga$10~Gyr old)
  are slightly overestimated. This may in fact be revealing the loss
  of temporal resolution beyond 7--8~Gyr old due to photometric
  uncertainties at faint magnitudes, as illustrated by the burst of
  Figure~\ref{fig:a4} being recovered with a lower amplitude and more
  spread out in time.
\end{itemize}

\begin{thebibliography}{}

\bibitem[\protect\citeauthoryear{Agertz, Teyssier, \&
    Moore}{2011}]{age11} Agertz O., Teyssier R., Moore B., 2011,
  MNRAS, 410, 1391

\bibitem[\protect\citeauthoryear{Aparicio \& Hidalgo}{2009}]{apa09}
  Aparicio A., Hidalgo S.~L., 2009, AJ, 138, 558

\bibitem[\protect\citeauthoryear{Aparicio \& Gallart}{2004}]{apa04}
  Aparicio A., Gallart C., 2004, AJ, 128, 1465

\bibitem[\protect\citeauthoryear{Barker et al.}{2011}]{bar11} Barker
  M.~K., Ferguson A.~M.~N., Cole A.~A., Ibata R., Irwin M., Lewis
  G.~F., Smecker-Hane T.~A., Tanvir N.~R., 2011, MNRAS, 410, 504

\bibitem[\protect\citeauthoryear{Barker et al.}{2007}]{bar07} Barker
  M.~K., Sarajedini A., Geisler D., Harding P., Schommer R., 2007, AJ,
  133, 1138

\bibitem[\protect\citeauthoryear{Bernard et al.}{2009}]{ber09} Bernard
  E.~J., et al., 2009, ApJ, 699, 1742

\bibitem[\protect\citeauthoryear{Bird, Kazantzidis, \&
    Weinberg}{2011}]{bir11} Bird J.~C., Kazantzidis S., Weinberg
  D.~H., 2011, MNRAS, {\it submitted} (astro-ph/1104.0933)

\bibitem[\protect\citeauthoryear{Braun et al.}{2009}]{bra09} Braun R.,
  Thilker D.~A., Walterbos R.~A.~M., Corbelli E., 2009, ApJ, 695, 937

\bibitem[\protect\citeauthoryear{Brown et al.}{2004}]{bro04} Brown
  T.~M., Ferguson H.~C., Smith E., Kimble R.~A., Sweigart A.~V.,
  Renzini A., Rich R.~M., 2004, AJ, 127, 2738

\bibitem[\protect\citeauthoryear{Brown et al.}{2006}]{bro06} Brown
  T.~M., Smith E., Ferguson H.~C., Rich R.~M., Guhathakurta P.,
  Renzini A., Sweigart A.~V., Kimble R.~A., 2006, ApJ, 652, 323

\bibitem[\protect\citeauthoryear{Cole et al.}{2007}]{col07} Cole
  A.~A., et al., 2007, ApJ, 659, L17

\bibitem[\protect\citeauthoryear{Collins et al.}{2011}]{col11} Collins
  M.~L.~M., et al., 2011, MNRAS, 413, 1548

\bibitem[\protect\citeauthoryear{Corbelli \& Schneider}{1997}]{cor97}
  Corbelli E., Schneider S.~E., 1997, ApJ, 479, 244

\bibitem[\protect\citeauthoryear{Courteau et al.}{2011}]{cou11}
  Courteau S., Widrow L.~M., McDonald M., Guhathakurta P., Gilbert
  K.~M., Zhu Y., Beaton R.~L., Majewski S.~R., 2011, ApJ, 739, 20

\bibitem[\protect\citeauthoryear{Cowie et al.}{1996}]{cow96} Cowie
  L.~L., Songaila A., Hu E.~M., Cohen J.~G., 1996, AJ, 112, 839

\bibitem[\protect\citeauthoryear{Cuillandre et al.}{2001}]{cui01}
  Cuillandre J.-C., Lequeux J., Allen R.~J., Mellier Y., Bertin E.,
  2001, ApJ, 554, 190

\bibitem[\protect\citeauthoryear{Dalcanton et al.}{2009}]{dalc09}
  Dalcanton J.~J., et al., 2009, ApJS, 183, 67

\bibitem[\protect\citeauthoryear{Di Matteo et al.}{2008}]{dim08} Di
  Matteo P., Bournaud F., Martig M., Combes F., Melchior A.-L.,
  Semelin B., 2008, A\&A, 492, 31

\bibitem[\protect\citeauthoryear{Duquennoy \& Mayor}{1991}]{duq91}
  Duquennoy A., Mayor M., 1991, A\&A, 248, 485

\bibitem[\protect\citeauthoryear{Fall \& Efstathiou}{1980}]{fal80}
  Fall S.~M., Efstathiou G., 1980, MNRAS, 193, 189

\bibitem[\protect\citeauthoryear{Fan, de Grijs, \& Zhou}{2010}]{fan10}
  Fan Z., de Grijs R., Zhou X., 2010, ApJ, 725, 200

\bibitem[\protect\citeauthoryear{Fardal et al.}{2007}]{fard07} Fardal
  M.~A., Guhathakurta P., Babul A., McConnachie A.~W., 2007, MNRAS,
  380, 15

\bibitem[\protect\citeauthoryear{Faria et al.}{2007}]{far07} Faria D.,
  Johnson R.~A., Ferguson A.~M.~N., Irwin M.~J., Ibata R.~A., Johnston
  K.~V., Lewis G.~F., Tanvir N.~R., 2007, AJ, 133, 1275

\bibitem[\protect\citeauthoryear{Ferguson et al.}{2002}]{fer02}
  Ferguson A.~M.~N., Irwin M.~J., Ibata R.~A., Lewis G.~F., Tanvir
  N.~R., 2002, AJ, 124, 1452

\bibitem[\protect\citeauthoryear{Ferguson et al.}{2005}]{fer05}
  Ferguson A.~M.~N., Johnson R.~A., Faria D.~C., Irwin M.~J., Ibata
  R.~A., Johnston K.~V., Lewis G.~F., Tanvir N.~R., 2005, ApJ, 622,
  L109

\bibitem[\protect\citeauthoryear{Ferguson \& Johnson}{2001}]{fer01}
  Ferguson A.~M.~N., Johnson R.~A., 2001, ApJ, 559, L13

\bibitem[\protect\citeauthoryear{Gallart, Zoccali, \&
    Aparicio}{2005}]{gal05} Gallart C., Zoccali M., Aparicio A., 2005,
  ARA\&A, 43, 387

\bibitem[\protect\citeauthoryear{Gallazzi et al.}{2008}]{gal08}
  Gallazzi A., Brinchmann J., Charlot S., White S.~D.~M., 2008, MNRAS,
  383, 1439

\bibitem[\protect\citeauthoryear{Galleti, Bellazzini, \&
    Ferraro}{2004}]{gal04} Galleti S., Bellazzini M., Ferraro F.~R.,
  2004, A\&A, 423, 925

\bibitem[\protect\citeauthoryear{Girardi et al.}{2000}]{gir00} Girardi
  L., Bressan A., Bertelli G., Chiosi C., 2000, A\&AS, 141, 371

\bibitem[\protect\citeauthoryear{Girardi \& Salaris}{2001}]{gir01}
  Girardi L., Salaris M., 2001, MNRAS, 323, 109

\bibitem[\protect\citeauthoryear{Gogarten et al.}{2010}]{gog10}
  Gogarten S.~M., et al., 2010, ApJ, 712, 858

\bibitem[\protect\citeauthoryear{Governato et al.}{2009}]{gov09}
  Governato F., et al., 2009, MNRAS, 398, 312

\bibitem[\protect\citeauthoryear{Grevesse \& Noels}{1993}]{gre93}
  Grevesse N., Noels A., 1993, in Origin and Evolution of the
  Elements, 15

\bibitem[\protect\citeauthoryear{Guedes et al.}{2011}]{gue11} Guedes
  J., Callegari S., Madau P., Mayer L., 2011, ApJ, 742, 76

\bibitem[\protect\citeauthoryear{Hammer et al.}{2010}]{ham10} Hammer
  F., Yang Y.~B., Wang J.~L., Puech M., Flores H., Fouquet S., 2010,
  ApJ, 725, 542

\bibitem[\protect\citeauthoryear{Haywood}{2008}]{haywood08} Haywood
  M., 2008, MNRAS, 388, 1175

\bibitem[\protect\citeauthoryear{Hidalgo et al.}{2011}]{hid11} Hidalgo
  S.~L., et al., 2011, ApJ, 730, 14

\bibitem[\protect\citeauthoryear{Holland}{1998}]{hol98} Holland S.,
  1998, AJ, 115, 1916

\bibitem[\protect\citeauthoryear{Horne \& Baliunas}{1986}]{hor86}
  Horne J.~H., Baliunas S.~L., 1986, ApJ, 302, 757

\bibitem[\protect\citeauthoryear{Ibata et al.}{2001}]{iba01} Ibata R.,
  Irwin M., Lewis G., Ferguson A.~M.~N., Tanvir N., 2001, Natur, 412,
  49

\bibitem[\protect\citeauthoryear{Ibata et al.}{2005}]{iba05} Ibata R.,
  Chapman S., Ferguson A.~M.~N., Lewis G., Irwin M., Tanvir N., 2005,
  ApJ, 634, 287

\bibitem[\protect\citeauthoryear{Irwin et al.}{2005}]{irw05} Irwin
  M.~J., Ferguson A.~M.~N., Ibata R.~A., Lewis G.~F., Tanvir N.~R.,
  2005, ApJ, 628, L105

\bibitem[\protect\citeauthoryear{Jarosik et al.}{2011}]{jar11} Jarosik
  N., et al., 2011, ApJS, 192, 14

\bibitem[\protect\citeauthoryear{Kennicutt et al.}{1987}]{ken87}
  Kennicutt R.~C., Jr., Roettiger K.~A., Keel W.~C., van der Hulst
  J.~M., Hummel E., 1987, AJ, 93, 1011

\bibitem[\protect\citeauthoryear{Kroupa}{2002}]{kro02} Kroupa P.,
  2002, Sci, 295, 82

\bibitem[\protect\citeauthoryear{Layden et al.}{1999}]{lay99} Layden
  A.~C., Ritter L.~A., Welch D.~L., Webb T.~M.~A., 1999, AJ, 117, 1313

\bibitem[\protect\citeauthoryear{Ma, Peng, \& Gu}{1997}]{ma97} Ma J.,
  Peng Q.-H., Gu Q.-S., 1997, ApJ, 490, L51

\bibitem[\protect\citeauthoryear{McConnachie et al.}{2009}]{mcc09}
  McConnachie A.~W., et al., 2009, Natur, 461, 66

\bibitem[\protect\citeauthoryear{McConnachie et al.}{2010}]{mcc10}
  McConnachie A.~W., Ferguson A.~M.~N., Irwin M.~J., Dubinski J.,
  Widrow L.~M., Dotter A., Ibata R., Lewis G.~F., 2010, ApJ, 723, 1038

\bibitem[\protect\citeauthoryear{Minchev et al.}{2011}]{min11} Minchev
  I., Famaey B., Combes F., Di Matteo P., Mouhcine M., Wozniak H.,
  2011, A\&A, 527, A147

\bibitem[\protect\citeauthoryear{Monelli et al.}{2010}]{mon10} Monelli
  M., et al., 2010, ApJ, 720, 1225

\bibitem[\protect\citeauthoryear{Mu{\~n}oz-Mateos et
    al.}{2011}]{mun11} Mu{\~n}oz-Mateos J.~C., Boissier S., Gil de Paz
  A., Zamorano J., Kennicutt R.~C., Jr., Moustakas J., Prantzos N.,
  Gallego J., 2011, ApJ, 731, 10

\bibitem[\protect\citeauthoryear{Newton \& Emerson}{1977}]{new77}
  Newton K., Emerson D.~T., 1977, MNRAS, 181, 573

\bibitem[\protect\citeauthoryear{Pietrinferni et al.}{2004}]{pie04}
  Pietrinferni A., Cassisi S., Salaris M., Castelli F., 2004, ApJ,
  612, 168

\bibitem[\protect\citeauthoryear{Putman et al.}{2009}]{put09} Putman
  M.~E., et al., 2009, ApJ, 703, 1486

\bibitem[\protect\citeauthoryear{Quillen et al.}{2009}]{qui09} Quillen
  A.~C., Minchev I., Bland-Hawthorn J., Haywood M., 2009, MNRAS, 397,
  1599

\bibitem[\protect\citeauthoryear{Reid \& Gizis}{1997}]{rei97} Reid
  I.~N., Gizis J.~E., 1997, AJ, 113, 2246

\bibitem[\protect\citeauthoryear{Reitzel, Guhathakurta, \&
    Rich}{2004}]{rei04} Reitzel D.~B., Guhathakurta P., Rich R.~M.,
  2004, AJ, 127, 2133

\bibitem[\protect\citeauthoryear{Rich et al.}{2004}]{ric04} Rich
  R.~M., Reitzel D.~B., Guhathakurta P., Gebhardt K., Ho L.~C., 2004,
  AJ, 127, 2139

\bibitem[\protect\citeauthoryear{Richardson}{2010}]{ric10} Richardson
  J.~C., 2010, PhD Thesis, University of Edinburgh

\bibitem[\protect\citeauthoryear{Richardson et al.}{2008}]{ric08}
  Richardson J.~C., et al., 2008, AJ, 135, 1998

\bibitem[\protect\citeauthoryear{Ro{\v s}kar et al.}{2008a}]{ros08a}
  Ro{\v s}kar R., Debattista V.~P., Quinn T.~R., Stinson G.~S.,
  Wadsley J., 2008a, ApJ, 684, L79

\bibitem[\protect\citeauthoryear{Ro{\v s}kar et al.}{2008b}]{ros08b}
  Ro{\v s}kar R., Debattista V.~P., Stinson G.~S., Quinn T.~R.,
  Kaufmann T., Wadsley J., 2008b, ApJ, 675, L65

\bibitem[\protect\citeauthoryear{Schlegel, Finkbeiner, \&
    Davis}{1998}]{sch98} Schlegel D.~J., Finkbeiner D.~P., Davis M.,
  1998, ApJ, 500, 525

\bibitem[\protect\citeauthoryear{Sch{\"o}nrich \&
    Binney}{2009}]{sch09} Sch{\"o}nrich R., Binney J., 2009, MNRAS,
  396, 203

\bibitem[\protect\citeauthoryear{Sellwood \& Binney}{2002}]{sel02}
  Sellwood J.~A., Binney J.~J., 2002, MNRAS, 336, 785

\bibitem[\protect\citeauthoryear{Sirianni et al.}{2005}]{sir05}
  Sirianni M., et al., 2005, PASP, 117, 1049

\bibitem[\protect\citeauthoryear{Skillman et al.}{2003}]{ski03}
  Skillman E.~D., Tolstoy E., Cole A.~A., Dolphin A.~E., Saha A.,
  Gallagher J.~S., Dohm-Palmer R.~C., Mateo M., 2003, ApJ, 596, 253

\bibitem[\protect\citeauthoryear{Sommer-Larsen, G{\"o}tz, \&
    Portinari}{2003}] {som03} Sommer-Larsen J., G{\"o}tz M., Portinari
  L., 2003, ApJ, 596, 47

\bibitem[\protect\citeauthoryear{Stetson}{1994}]{ste94} Stetson P.~B.,
  1994, PASP, 106, 250

\bibitem[\protect\citeauthoryear{Stetson}{1996}]{ste96} Stetson P.~B.,
  1996, PASP, 108, 851

\bibitem[\protect\citeauthoryear{Tabatabaei \&
    Berkhuijsen}{2010}]{tab10} Tabatabaei F.~S., Berkhuijsen E.~M.,
  2010, A\&A, 517, A77

\bibitem[\protect\citeauthoryear{Teyssier, Chapon, \&
    Bournaud}{2010}]{tey10} Teyssier R., Chapon D., Bournaud F., 2010,
  ApJ, 720, L149

\bibitem[\protect\citeauthoryear{van der Wel et al.}{2011}]{van11} van
  der Wel A., et al., 2011, ApJ, 730, 38

\bibitem[\protect\citeauthoryear{VandenBerg, Bergbusch, \&
    Dowler}{2006}]{van06} VandenBerg D.~A., Bergbusch P.~A., Dowler
  P.~D., 2006, ApJS, 162, 375

\bibitem[\protect\citeauthoryear{Williams}{2002}]{wil02} Williams
  B.~F., 2002, MNRAS, 331, 293

\bibitem[\protect\citeauthoryear{Williams et al.}{2009a}]{wil09a}
  Williams B.~F., et al., 2009a, AJ, 137, 419

\bibitem[\protect\citeauthoryear{Williams et al.}{2009b}]{wil09b}
  Williams B.~F., Dalcanton J.~J., Dolphin A.~E., Holtzman J.,
  Sarajedini A., 2009b, ApJ, 695, L15

\bibitem[\protect\citeauthoryear{Williams et al.}{2010}]{wil10}
  Williams B.~F., et al., 2010, ApJ, 709, 135

\bibitem[\protect\citeauthoryear{Wong et al.}{2011}]{won11} Wong
  K.~C., et al., 2011, ApJ, 728, 119

\bibitem[\protect\citeauthoryear{Wright et al.}{1972}]{wri72} Wright
  M.~C.~H., Warner P.~J., Baldwin J.~E., 1972, MNRAS, 155, 337

\bibitem[\protect\citeauthoryear{Wyse}{2008}]{wys08} Wyse R.~F.~G.,
  2008, ASPC, 399, 445

\end{thebibliography}
\end{document}